\documentclass[runningheads]{llncs}
\def \vmcai {}
\input{preamble.tex}

\begin{document}

% \nopagecolor

\title{Sound Symbolic Execution via Abstract Interpretation and its Application to Security} 
\titlerunning{Sound Symbolic Execution via Abstract Interpretation}
\iffinal{
  \author{Ignacio Tiraboschi\inst{2} \and
    Tamara Rezk\inst{1} \and
    Xavier Rival\inst{2}}
  \authorrunning{Tiraboschi et al.}
  \institute{
    % INRIA Sophia Antipolis, France
    INRIA, Université Côte d'Azur, Sophia Antipolis, France
    % \email{name.surname@inria.fr}
    \and
    INRIA Paris, DI ENS, Ecole normale supérieure, Université PSL, CNRS\\
    \email{name.surname@inria.fr}}
}{
  \author{}
  \institute{}
}
\maketitle
\begin{abstract}
  Symbolic execution is a program analysis technique commonly utilized
  to determine whether programs violate properties and, in case
  violations are found, to generate inputs that can trigger them.
  Used in the context of security properties such as
  noninterference, symbolic execution is  precise when looking for
  counter-example pairs of traces when insecure information flows are
  found, however it is sound only up to a bound thus it does not allow
  to prove the correctness of programs with executions beyond the given bound.
  By contrast, abstract interpretation-based static analysis guarantees
  soundness but generally lacks the ability to provide counter-example pairs
  of traces.

  In this paper, we propose to weave both to obtain the best of two worlds.
  We demonstrate this with a series of static analyses, including a  
  static analysis called \DSym aimed
  at verifying noninterference. \DSym provides both semantically sound
  results and the ability to derive counter-example pairs of traces up
  to a bound.
  It relies on a combination of symbolic execution and abstract domains
  inspired by the well known notion of reduced product.
  We formalize \DSym and prove its soundness as well as its relative
  precision up to a bound.
  We also provide a prototype implementation of \DSym and evaluate it on
  a sample of challenging examples.
  \iffinal{}{
  }
\end{abstract}
\begin{center}
  \includegraphics[scale=0.14]{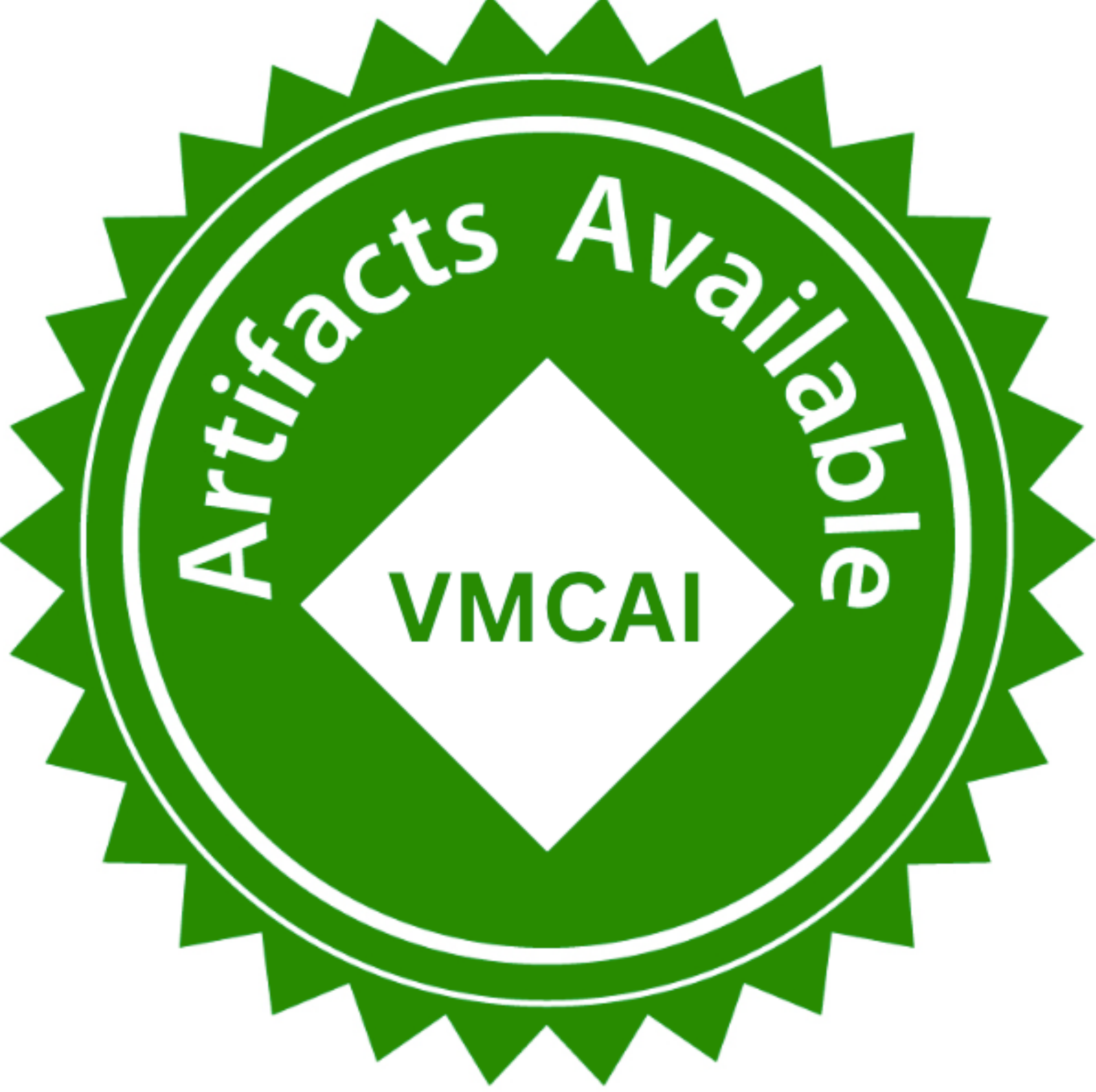} 
  \includegraphics[scale=0.14]{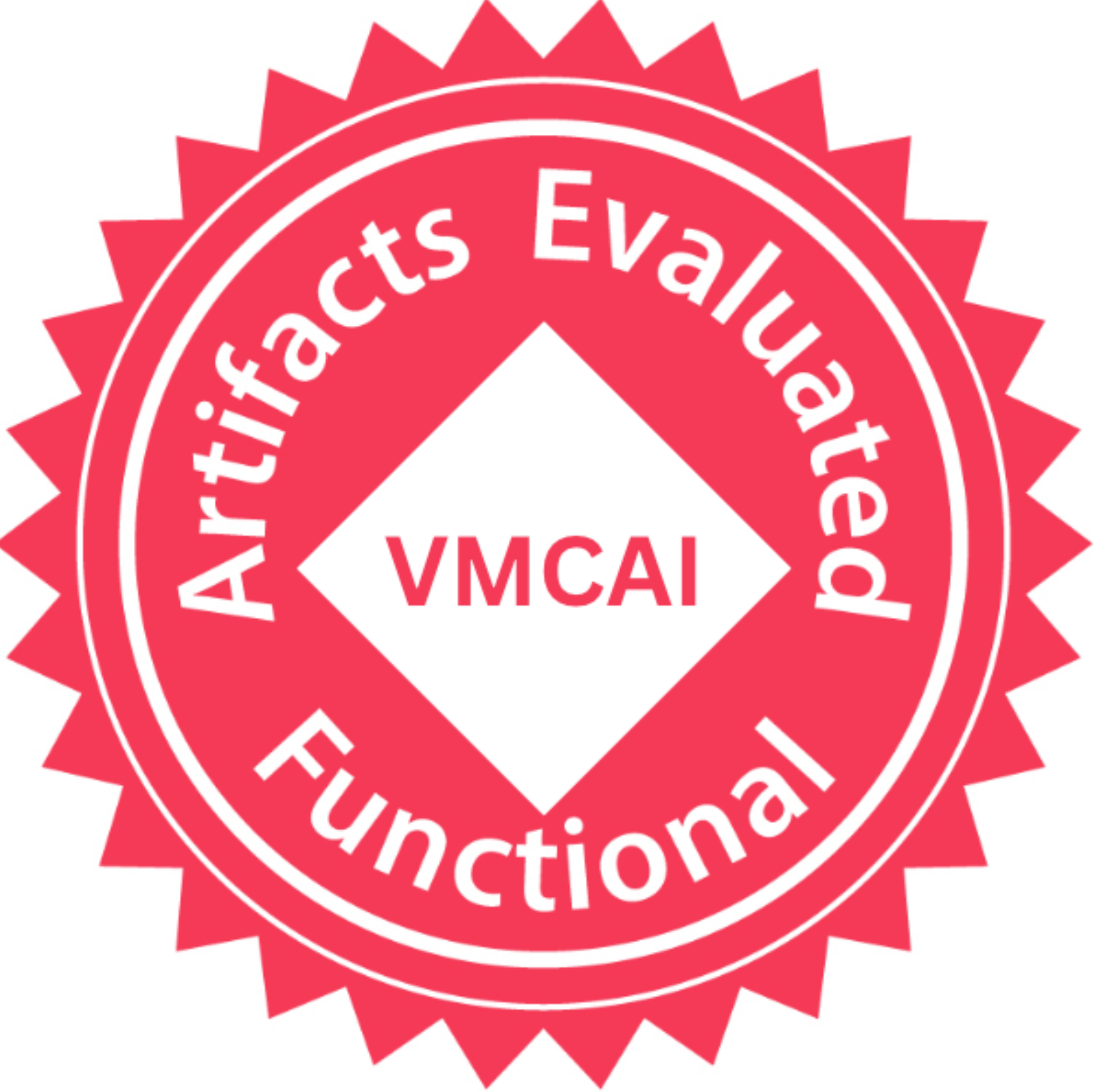}
\end{center}

\section{Introduction}
% Figure of different analysis and how they relate.
\begin{figure}[t]
\centering
\begin{tikzpicture}
  % Nodes
  \node (SE)    at (-3.05,0) {SE~\cite{bel:acm:75}};
  \node (RSE)   at (-3.05,-1) {RSE~\cite{mbc:forte:12}};
  \node (SSE)   at (0,0) {\SoundSE{\scriptsize(Sec.~\ref{s:4:ssym})}};
  \node (SRSE)  at (0,-1) {\SoundRSE{\scriptsize(Sec.~\ref{s:6:dsym})}};
  \node (RSSE)  at (3.5,0) {\RedSoundSE{\scriptsize(Sec.~\ref{s:5:sabs})}};
  \node (RSRSE) at (3.5,-1) {\RedSoundRSE{\scriptsize(Sec.~\ref{s:7:sdep})}};
  \node         at (-5.2,0) {\scriptsize Single trace analyses:};
  \node         at (-5.2,-1) {\scriptsize Relational analyses:};
  % \node[text width=3cm, align=center, above=of SSE,  anchor=east, rotate=-45, xshift=0.8cm, yshift=-0.3cm] {\scriptsize Sound analyses:};
  % \node[text width=3cm, align=center, above=of RSSE, anchor=east, rotate=-45, xshift=0.5cm] {\scriptsize Reduced product and sound analyses:};
  \node[text width=3cm, align=center, above=of SSE, anchor=south, yshift=-0.5cm] {\scriptsize Sound analyses:};
  \node[text width=3cm, align=center, above=of RSSE, anchor=south, yshift=-0.5cm] {\scriptsize Reduced product and sound analyses:};
  % Semantic arrows
  % \node at (0,0.35) {$(\sem)$};
  \node at (0,0.4) {$(\sse)$};
  \node at (3.5,0.4) {$(\rsse)$};
  % \node at (0,-1.35) {$(\sem)$};
  \node at (0,-1.4) {$(\dse)$};
  \node at (3.5,-1.4) {$(\rdse)$};
  % Arrows
  \draw[->, draw=blue, dashed] (SE) -- (SSE);
  \draw[->, draw=blue, dashed] (SSE) -- (RSSE);
  \draw[->, draw=blue, dashed] (RSE) -- (SRSE);
  \draw[->, draw=blue, dashed] (SRSE) -- (RSRSE);
  \draw[->, draw=red, dashed] (SE) -- (RSE);
  \draw[->, draw=red, dashed] (SSE) -- (SRSE);
  \draw[->, draw=red, dashed] (RSSE) -- (RSRSE);
\end{tikzpicture}
\caption{Relation between different SE analyses. SE~\cite{bel:acm:75} is conventional symbolic
execution and RSE~\cite{mbc:forte:12,pkc:icse:16} is its extension to relational properties.
Except for RSE with invariants~\cite{fcg:ppdp:19}, SE and RSE are unsound in general. 
The rest of the analyses are sound and are our contributions:
\SoundSE and \SoundRSE do not use abstract interpretation
whereas \RedSoundSE and \RedSoundRSE can be combined with different
abstract domains.
A red dashed line 
represents  a dependency: a relational analysis  depends on a single trace analysis. 
A blue dashed line represents an enhancement of the analysis.}
\label{fig:symbolic_exec}
\end{figure}

Security properties are notoriously hard to verify.
In particular, many security properties are not single-execution properties but
hyperproperties~\cite{cs:hyper:08} (also referred to as relational
properties), which means that refuting them \modification{sometimes} requires \emph{several}
executions traces to be provided as a counter-example.
In particular, noninterference~\cite{gm:sp:82} states that high clearance
information should not impact the observation of low clearance users in any
execution of the program.
It has been the subject of many verification method proposals and
tools (e.g.~\cite{sm:ieee:03,bar:csf:04,ss:ieee:05,fpr:ccs:11,fjrs:tgc:15,brr:post:16,fcg:ppdp:19,anst:popl:17,nbfrrs:www:18}).

\emph{Symbolic execution}~\cite{bel:acm:75,king:acm:76} (SE) is typically
used to find property violations, and can be applied for policies like
noninterference provided some adaptation for relational properties.
SE boils down to an execution where variables initially hold symbolic values
and get updated with expressions of these symbolic values whereas conditions
are evaluated into symbolic path guards. The analysis involves an external tool such
as an SMT solver that prunes infeasible paths and attempts to discharge
verification conditions on remaining ones.
SE attempts to exhaustively cover all executions paths, which is feasible
only up to a bound and quickly turns out costly in presence of unbounded
loops.

Conventional SE does not over-approximate executions after a fixed bound of
iterations. This implies that \emph{soundness} is lost when the program
exceeds the exploration bound.
% , thus soundness is guaranteed only when the program does not exceed the exploration bound.
Soundness ensures that, when the analysis concludes that the property of
interest holds, the concrete semantics of the analyzed program is guaranteed
to satisfy it.
Since there is no over-approximation, when the property is violated by traces shorter than
the exploration bound, tools like SMT solvers can provide instances for
the symbolic values and enable the reconstruction of counter-example
traces.
This is of particular importance to security in order to confirm
security violations.
We refer to such counter-examples as {\bf refutation models}.

The adaptation of SE to handle relational
properties~\cite{mbc:forte:12,pkc:icse:16}  requires to track several traces instead of just one.
In the following, we will call this adaptation \emph{relational symbolic execution} (or RSE).
Previous work~\cite{fcg:ppdp:19} has shown how to combine
RSE with loop invariants, provided by the developer, in order to recover
soundness at the cost of annotations and loss of precision when invariants
are not strong enough. 

%By contrast, 
\emph{Abstract interpretation based static
  analyses}~\cite{cc:popl:77} (AI) rely on an abstraction defined as a logical
approximation relation between concrete behaviors and abstract predicates
and produce sound over-approximations of program semantics at the cost of
completeness.
% Soundness ensures that, when the analysis concludes that the property of
% interest holds, the concrete semantics of the analyzed program is guaranteed
% to satisfy it.
However, the over-approximation entails that the analysis may
fail to conclude positively even when analyzing correct programs.
Moreover, most static analysis implementations lack the ability to
synthesize counter-example traces.

In this paper, we formalize a combined analysis technique, which aims at
bringing together advantages of both symbolic execution and
abstract interpretation, in a security setup.
\modification{We first show how to over-approximate SE in order to keep soundness, and call this analysis \SoundSE (see Figure~\ref{fig:symbolic_exec}).}
We use \SoundSE to show the combination for
conventional SE and different abstract domains, calling the resulting analyses \RedSoundSE.
Our analysis for relational properties is called \DSym and targets noninterference.
It borrows path exploration from relational symbolic execution, parameterized by \RedSoundSE, and relies on abstract
interpretation based static analysis to report a sound result for all
programs.
Abstraction enables the early pruning of infeasible paths and the
computation of sound over-approximations for program behaviors when the
exploration bound is exhausted.
To achieve this, \DSym automatically injects loop invariants computed
by abstract domains into a relational store.
Not only dependence analysis results can be used to fill security related
information where the symbolic execution cannot explore paths fully but
also (e.g., numerical) state abstraction information allows to improve
the symbolic information extracted from the dependency analysis.
Moreover, our analysis allows switching between  different abstractions, and tuning specific
settings, e.g., loop unroll depth (depth up to which SE is kept precise), which allows the user to
change the balance between cost and precision.
To summarize, we propose symbolic execution based verification methods that are
sound and  precise, providing refutation models up to a bound. Our contributions,
illustrated in Figure~\ref{fig:symbolic_exec}, are the following:
\begin{compactenum}
\item \SoundSE and \RedSoundSE: We define a sound SE analysis, and we integrate numerical abstract
  domains into it to prune reachable paths. As a result, we make SE~\cite{bel:acm:75} sound while
  keeping the ability of the analysis to find counter-examples.
\item \SoundRSE and \RedSoundRSE: We define a sound relational SE, and we combine it with dependence
  analysis~\cite{anst:popl:17} to enhance the precision of the latter while preserving soundness.
\item We prototype \DSym \modification{together with \RedSoundSE} in OCaml and show, using a series of challenging examples, that it is able to both soundly decide noninterference for secure programs and  synthesize counter-examples of a size up to a given bound for insecure ones.
\end{compactenum}
The structure of the paper is as follows.
\sref{3:lang} defines a basic language and the noninterference notion used throughout the
rest of the paper.
\sref{2:over} provides an overview on already defined analyses and highlights the main principles of \DSym. % compared to existing methods.
\sref{4:ssym} defines \SoundSE, a sound single trace symbolic execution that serves
as a basis for \DSym and \sref{5:sabs} presents \RedSoundSE, a new combination of \SSym
with state abstraction. % (e.g., value)
%\sref{6:dsym} formalizes the symbolic execution part of \DSym and
\sref{6:dsym} presents \SoundRSE and
\sref{7:sdep} extends it with a dependence abstraction to obtain \DSym.
\sref{8:eval} evaluates our framework on small but challenging examples.
Finally, \sref{9:rwc} discusses related work and \sref{10:conc} concludes.
The appendix contains all rules of analyses in the paper.

\section{Language and noninterference security notion}
\label{sec:lang} \label{s:3:lang}

%In this paper, we formalize our analysis based on the simple imperative
%language defined by the grammar below.
%We let $\Val$ denote the set of values (e.g., for the sake of simplicity, we
%may consider mathematical integers) and $\Var$ denote a countable set of
%program variables.
%As usual, an expression $\prexpr$ is either a value or the reading of a
%variable, or a binary operator applied to two expressions; it evaluates
%into a value.

In this section, we introduce the language and security notion for which we formalize our analyses.
We let $\Val$ and $\Var$ be the set of
values and program variables respectively, and $\oplus$, $\olessthan$ be binary operators.
A boolean expression $\prbcond$ is a comparison operator $\olessthan$ applied to two
expressions and evaluates to a boolean value $\sBool = \{ \true, \false \}$.
A statement $\prstat$ is either a skip,
an assignment, a condition, or a loop.
Finally, a command $\prcom$ is a finite sequence of statements.
A program $\prog$ is a pair $(\prcom,\prlowvars)$ made of
a command $\prcom$ (the body of the program), and a set of low variables
$\prlowvars \subseteq \Var$, hence publicly observable (the other
variables occurring in the program are high).
\[
\begin{array}{rclcrcl}
  \prexpr  & ::= & v\ (v \in \Val)\
  |\ x\ (x \in \Var)\
  |\ \prexpr \oplus \prexpr
  & \qquad &
  \prbcond & ::= & \prexpr \olessthan \prexpr
  \\
  \prstat  & ::= & \stmtskip\
  |\ x := \prexpr\
  |\ \stmtif{\prbcond}{\prcom}{\prcom}\
  |\ \stmtwhile{\prbcond}{\prcom}
  & \qquad &
  \prcom  & ::= & \prstat\
  |\ \prstat; \prcom
  \\
\end{array}
\]

\paragraph{Semantics.}
Given a program $(\prcom,\prlowvars)$, a \emph{state} is a pair $(\prcom, \mem)$, where
$\prcom$ is a command and  $\mem$ is  a function from $\Var$ to $\Val$, namely a \emph{store}.
In particular, a state of the form $(\kwskip, \mem)$ is final.
% (i.e.,corresponds to an execution that has terminated).
We write $\Mem$ and $\Conf$ for the
set of stores and states respectively. %NOT SURE WE NEED TO KEEP \Conf for vmcai
%When enumerating explicitly the contents of a store, we use the
We use  $[\varx \mapsto x, \vary \mapsto y, \ldots]$ to explicitly enumerate a store's contents, where
$x, y, \ldots$ are concrete values. %NO TIME FOR THIS NOW BUT WE LET VALUES RANGE OVER V NOT X ....
Let $(\sem) \subseteq \Conf \times \Conf$ denote the small step
operational semantics (which is standard) and $\sems$ be its reflexive transitive closure.
%A trace is defined as an element $t$ of $\Trc \triangleq \Conf \times
%\Conf$ and describes an execution with its initial and final states.

\paragraph{Noninterference.}
%For all $\mem_0,\mem_1$, and  $\prlowvars$,
Let $\loweq$ be
the set equality of stores restricted to low variables in $\prlowvars$.
In the rest of the paper, we focus on termination-insensitive noninterference:
%We let $\varx \mapsto x =_{\vary} \varx \mapsto x'$ be $x=
\begin{definition}[Termination-insensitive noninterference]
  \label{d:1:ni}
  A program $(\prcom,\prlowvars)$  is \emph{termination-insensitive
    noninterferent}, written as $\niti_\prog$, if and only if,
  for all stores $\mem_0, \mem_1, \mem'_0, \mem'_1 \in \Mem$,
  $
  \mem_0 \loweq \mem_1
  \logand (\prcom,\mem_0) \sems (\kwskip, \mem_0')
  \logand (\prcom,\mem_1) \sems (\kwskip, \mem_1')
  \Longrightarrow
  \mem_0' \loweq \mem_1'
  $.
\end{definition}

\section{Overview}
\label{sec:over} \label{s:2:over}
In this section, we demonstrate the principle of the combination of
symbolic execution and abstraction performed by \DSym so as to overcome
the limitation of these two approaches taken separately.
As in the rest of the paper, we focus on noninterference (NI), although
the same principle would apply to other security properties as well.

% \paragraph{Noninterference.}
% NI captures the idea that a program should not let low clearance (or for
% short, low) variables (i.e., that can be seen by anyone) convey any
% information about the values of private variables (that hold secret
% information, aimed to be accessed only by authorized users).
% It is well known to be a 2-safety property \cite{cs:hyper:08}:
% observing individual traces provides no information about noninterference,
% and reasoning about it requires to look at pairs of executions.
% More formally, a program $\prog$ is noninterferent if and only if, for any
% \modification{two concrete stores (mappings from variables to values)} $\mem_0, \mem_1$ that are initially equal in all low
% variables of $\prog$, if $\mem_0', \mem_1'$ are the final stores after
% executing $\prog$ from $\mem_0$ and $\mem_1$ respectively, then $\mem_0'$
% and $\mem_1'$ should also be equal over all low variables.
% In other words, a potential attacker may not guess, just by observing low
% variables, any information giving a clue about the value of private variables.
% \modification{We use notation $\mem_0 \loweq \mem_1$ to indicate that all low variables in $\mem_0$ and $\mem_1$
% are equal, and call this relation \emph{low-equality}.}

\paragraph{Examples.}
We consider the programs displayed in \fref{1:ex}.
Essentially, programs (a) and (b) are secure  with respect
to the noninterference policy, where \lstinline|priv| is high and all
other variables are low, whereas (c) is not secure.
\begin{figure}[t]
  \resizebox{\textwidth}{!}{
    \subfigure[Secure program\label{lst:label2}]{ \label{fp:a}
\begin{lstlisting}[style=c,numbers=left]
if (priv > 0)
  y = 5;
else
  y = 5;
\end{lstlisting}}
    \qquad \quad
    \subfigure[Secure program\label{lst:label1}]{ \label{fp:b} \label{f:fp:b}
\begin{lstlisting}[style=c,numbers=left]
while (i < z) {
  i = i + 1;
  priv = priv + 5;
}

\end{lstlisting}}
    \qquad \quad
    \subfigure[Insecure program\label{lst:label3}]{ \label{fp:c}
\begin{lstlisting}[style=c,numbers=left]
while (i > priv) {
  i = i + 1;
  priv = priv + 2;
}


\end{lstlisting}}
    \qquad \quad
    \subfigure[A secure program requiring a numerical domain.]{ \label{f:3:ex}
\begin{lstlisting}[style=c,numbers=left,xleftmargin=0.38\textwidth]
if (priv < 0) priv = 0;
while (i < 10){
  i += 1; priv += 2;
}
if (priv >= 0) y += 1;
else y = 0;

\end{lstlisting}}
  }
  \caption{Example programs. All variables are of type {\ttfamily int}, where variable is $\varpriv$ is secure.}
  \label{f:1:ex}
\end{figure}

% In program \ref{fp:a}, the only low variable is variable \lstinline|y|,
% which gets assigned the value 5 whatever the execution path, so it provides
% no clue about the private variables and thus the program is secure.
In program~\ref{fp:a}, variable \vary\ gets assigned 5 independently of \varpriv, therefore the
program is secure.
For Program~\ref{fp:b}, let $\mem_0, \mem_1$ be two stores such that $\mem_0 \loweq \mem_1$.
Since $\mem_0$ and $\mem_1$ are low-equal executions cannot take different paths, and the loop will
be executed the same amount of times. Therefore, the program is secure.
Lastly, Program~\ref{fp:c} is insecure, meaning that it does not satisfy noninterference.
We need to provide a counter-example consisting of two
executions starting from low-equal stores $\mem_0, \mem_1$ such that the corresponding output stores
$\mem'_0, \mem'_1$ are not low-equal.
We consider the following stores: $\mem_0 = [ \vari \mapsto 0, \varpriv \mapsto 0 ]$, and
$\mem_1 = [ \vari \mapsto 0, \varpriv \mapsto -1 ]$.
Finally, calculated output stores are such that $\mem'_0(\vari) = 0 \not= \mem'_1(\vari) = 1$, thus
the program violates noninterference.

In the next paragraphs we study the result of verification methods
for these three programs.
% \erasable{
% (Table~\ref{f:2:verif} summarizes these results).
% }

\paragraph{Verification based on relational symbolic execution.}
A symbolic store, referred to as $\sstore$, maps variables to symbolic expressions of the initial
values of the variables.
To avoid confusion, we use an italic typewriter font for these symbolic
values while program variables appear in straight typewriter
font.
For instance, $\syvary$ denotes the initial value of \vary.
% \erasable{
% As an example, the symbolic execution of \lstinline{y = y + 1} results in
% the symbolic store $[\vary \mapsto \syvary+1]$.
% }
Relational symbolic execution describes pairs of executions using
symbolic conditions over the initial values of variables and pairs of
symbolic stores.
% \modification{which we refer to as \emph{relational symbolic store}.}
% \erasable{
% In such a symbolic store, to discriminate entities related to the two
% executions, we index with 0 and 1 each of the variables (symbolic or not).
% }
Symbolic stores are not enough to abstract executions, since they cannot express constraints.
\modification{Constraints are then provided by a \emph{symbolic path} $\spath$ that contextualizes
the store.
A pair $(\sstore, \spath)$ of a symbolic store and a symbolic path is referred to as a
\emph{symbolic precise store}.}
% \erasable{
\begin{table}[t]
  \begin{center}
    \begin{tabular}{ l c l l l }
      \toprule
      & Secure?
      & RSE
      & Dependence analysis $\quad$
      & \RedSoundRSE
      \\ \midrule
      Program \ref{fp:a}
      & Yes
      & \cmark\ Secure
      & \xmark\ False alarm
      & \cmark\ Secure
      \\ \midrule
      Program \ref{fp:b}
      & Yes
      & \xmark\ False alarm
      & \cmark\ Secure
      & \cmark\ Secure
      \\ \midrule
      Program \ref{fp:c}
      & No
      & \cmark\ Refutation model $\quad$
      & \cmark\ Alarm
      & \cmark\ Refutation model
      \\ \bottomrule
    \end{tabular}
  \end{center}
  \caption{Analysis results compared.
    Symbol \cmark\ (resp., \xmark\ ) denotes a semantically correct (resp.,
    incorrect) analysis outcome, with either a proof of security, a
    (possibly false) alarm, or a refutation model.}
  \label{f:2:verif}
  % \vspace{-1.5em}
\end{table}
% }

As an example, we consider Program \ref{fp:a}.
Relational symbolic execution uncovers four pairs of paths depending on
the sign of the initial values of \varpriv\ in both executions.
For instance, one of the diverging paths produces $\spath = (\syvarpriv_0 > 0
\logand \syvarpriv_1 \leq 0) \Longrightarrow ([ \vary_0 \mapsto
5, \ldots ], [\vary_1 \mapsto 5, \ldots ])$, where $\vary_0$ and
$\vary_1$ denote the program variable $\vary$ in both executions
and $\syvarpriv_0, \syvarpriv_1$ the initial symbolic values of \varpriv.
This symbolic precise store shows no information flow to \vary\ since any
SMT solver can prove $\vary_0 = \vary_1$.
The other three pairs of paths lead to a similar result, thus the program
is proved secure.

For Program \ref{fp:b}, the loop has an unbounded number of
iterations, but relational symbolic execution can only cover finitely
many unrollings of the loop.
This prevents RSE to prove that Program \ref{fp:b} is secure.
% \erasable{
% which prevents relational symbolic execution to compute such
% a set of pairs of symbolic paths.
% As a consequence, the symbolic execution requires a bound on the number
% of loop iterations that will be covered and stops beyond this bound.
% %gives up all precision beyond this bound.
% Therefore, program \ref{fp:b} cannot be proved secure.
% }

% The case of Program \ref{fp:c} is interesting as we remarked that this program
% is not secure.
For Program \ref{fp:c}, RSE will only explore the loop up to a bound.
Assuming the bound is one (any positive value would prove similar), it can determine that the
program does not satisfy NI by calculating a concrete trace that violates the property.
This counter-example trace is calculated by an SMT solver, for instance $\syvari_0 = \syvari_1 = 0$, $\syvarpriv_0 = 1$
and $\syvarpriv_1 = -1$ corresponds to the counter-example given previously.
% Since it contains a loop, relational symbolic execution will only explore
% the loop up to a bound.
% Let us assume that this bound is one (any strictly positive value would
% give a similar result).
% \erasable{
% Then the output below characterizes the case where the left execution
% does not enter the loop and the right one performs exactly one iteration:
% \[
% \left.
%   \begin{array}{ll}
%     & \syvari_0 \leq \syvarpriv_0 \\
%     \logand & \syvari_1 > \syvarpriv_1
%     \logand \syvari_1 + 1 \leq \syvarpriv_1 + 2 \\
%   \end{array}
% \right\}
% \Longrightarrow
% \left\{
%   \begin{array}{l}
%     \lbrack \vari_0 \mapsto \syvari_0,
%     \varpriv_0 \mapsto \syvarpriv_0 \rbrack, \\
%     \lbrack \vari_1 \mapsto \syvari_1 + 1,
%     \varpriv_1 \mapsto \syvarpriv_1 + 2 \rbrack \\
%   \end{array}
% \right.
% \]
% By adding the constraint that stores should be low-equal
% % executions should start from initial states that are low equal
% ($\syvari_0 = \syvari_1$), an SMT solver will
% produce a model (a concrete trace), for instance: $\syvari_0 = \syvari_1 = 0$, $\syvarpriv_0 = 1$
% and $\syvarpriv_1 = -1$, which corresponds to the
% counter-example previously given.
% % ?that we pointed out at the beginning of the section.
% }

\paragraph{Verification based on dependence abstraction.}
Many static analyses that work for noninterference rely on some form of
dependence abstraction as formalized in, e.g., \cite{anst:popl:17} or
\cite{hs:popl:06}.
We briefly summarize the abstraction of \cite{anst:popl:17}.
We assume an ordered set of security levels $\{ \Low, \High \}$ and
that each value fed into a program via an input variable is given a
security level.
% Then, a dependency, noted as $l \leadsto \varx$ where $l$ is a security
% level and $\varx$ is a variable, expresses that ``given two executions that
% agree on values up to security level $l$, then final states agree on \varx''.
\modification{%
A dependency, noted as $l \leadsto \varx$ with $l \in \{ \Low, \High \}$, expresses the agreement of
\varx\ in both executions when observing from level $l$.
This analysis, based on abstract interpretation, is \emph{sound}.
}
% \erasable{
% In that setup, dependency analysis computes dependency states made of such
% constraints.
% }

We now discuss the analysis of some programs in \fref{1:ex}.
% \erasable{
% We first consider program \ref{fp:a}.
% The initial dependency state should express that only \varpriv\ is not low,
% which means that it is $[ \Low \leadsto \{ \vary \} ;\ \High \leadsto
% \{ \vary, \varpriv \} ]$.
% Starting from this dependency state, the analysis discovers that the
% condition test may be influenced by a high variable, which induces an
% implicit information flow on all the assignments performed in the body
% of the condition statement.
% This leads to the dependency $\Low \leadsto \vary$ being dropped.
% As a consequence, the analysis concludes this program may be insecure.
% Note that this result is sound, but not complete, since the analysis
% does not guarantee the program is insecure either.
% }
\modification{%
For Program~\ref{fp:a},
the analysis determines that the assignments are conditioned by the value of \varpriv, which is
initially high.
Then, the dependency $\Low \leadsto \vary$ is dropped, indicating that \vary\ can potentially
disagree between executions.
}%
% \erasable{
% We now turn to program \ref{fp:b}.
% The initial dependence state is $[ \Low \leadsto \{ \vari, \varz \} ;\
% \High \leadsto \{ \vari, \varz, \varpriv \} ]$.
% Under this assumption, the loop condition is influenced only by \vari\
% and \varz\ which are of level \Low.
% Finally, the statement \lstinline{i = i + 1} preserves the fact that
% \vari\ is of level \Low\ which allows the analysis to converge.
% Therefore, all the initially assumed low variables remain so and the
% program is proved secure.
% }
%
\modification{%
In Program~\ref{fp:b}, the loop condition is only influenced by \vari\ and \varz, which are low.
Then, the assignment of low variables is not affected, and \vari\ and \varz\ remain low, allowing to
prove noninterference.
}%

% \erasable{
% Last, in the case of Program~\ref{fp:c}, as we observed that noninterference
% does not hold and dependency analysis is sound, we expect it to report an
% inconclusive answer.
% Indeed, the analysis discards the dependency $\Low \leadsto \vari$ due to
% the implicit flow at the loop condition point.
% }
\modification{%
Lastly, Program~\ref{fp:c} is not secure, and since dependence analysis is sound, the analysis
discards dependency $\Low \leadsto \vari$ based on the illicit flow of information.
}%

\paragraph{Combination of relational symbolic execution and dependence
  abstraction.}
As observed in Table~\ref{f:2:verif}, relational symbolic execution fails to handle precisely
program \ref{fp:b} whereas dependence abstraction fails to verify program
\ref{fp:a} and provides no counter-example for program \ref{fp:c}.
The purpose of \DSym is to
\modification{%
use both techniques in an alternating manner in order to increase precision and prune branches.
}%
% \erasable{
% take the best of both analyses into a unique analysis.
% }

To achieve this, \DSym borrows from relational symbolic execution the
precise analysis of assignment and condition commands, as well as the
unrolled iterates of loop commands.
In particular, the analysis of programs \ref{fp:a} and \ref{fp:c} is
carried out as shown above.
However, when the unrolling bound is reached, dependence analysis is used
% it uses dependence analysis
as a means to compute in finite time sound information about
any number of further loop iterations.
Indeed, when the dependence information proves that a loop induces no
dependency of a given low variable on any high variable, it is possible
to assume the equality of the variable in the symbolic store.
This new value may not be expressed precisely in terms of the initial
values, hence it may be approximated with a fresh symbol.
This occurs for variable \vari\ in program \ref{fp:b}.

\modification{As seen, \RedSoundRSE analyzes the first three examples of Figure~\ref{f:1:ex} precisely.}
% \erasable{
% As shown in Table~\ref{f:2:verif}, \DSym analyzes precisely the three
% examples of \fref{1:ex}.
% \TODO{Maybe remove table?}
% }
% \modification{
% The combination of dependences and SE is
% formalized in \sref{7:sdep}.
% }

\paragraph{Refinement of symbolic execution based on state abstraction.}
% We now observe that keeping sound symbolic execution precise requires more
% than just dependence information.
Program~\fref{3:ex}, previously not considered, cannot be proved NI by just using symbolic execution
and dependence analysis.
This program is secure since the assignment of \vari\ does
not depend on \varpriv, and \vary\ is conditioned by \varpriv\ which is always
positive after the loop.
% We also remark that the value of \varpriv\ at the loop exit is always
% positive, thus only the first branch of the condition may be taken,
% hence, there is no information flow to \vary\ either.

% \begin{figure}[t]
%   \begin{center}
%     \lstinputlisting[style=c,numbers=left,xleftmargin=0.38\textwidth]{prog-d.c}
%   \end{center}
%   \caption{A secure program requiring more than symbolic execution and
%     dependences.}
%   \label{f:3:ex}
% \end{figure}
As in Program~\ref{fp:b}, the loop causes the symbolic
execution to stop at the unrolling bound.
Dependence information allows to prove that there is no information
flow to $\vari$ and also that the value of $\vary$ at line 8 does not
depend on $\varpriv$.
However, the condition at line 9 depends on $\varpriv$, thus dependence
analysis will not prove that the assignments at lines 10 and 12 do not
leak information.
Symbolic execution does not succeed either as it lacks the ability to
reason over the value of $\varpriv$ at the loop exit.

Such information may be computed using a reachability static
analysis.
In particular, a classical static analysis based on the abstract domains
of intervals~\cite{cc:popl:77} computes ranges for all numeric variables
and concludes in this case that $\varpriv$ is positive, hence only the
true branch of the condition may be taken.
\modification{%
% We can enhance our analysis by integrating this abstract domains, in a product domain with a non
% relational SE.
% The analysis is then able to dynamically prune the paths that do not consider \varpriv\ positive.
Integrating non-relational abstract domains allows the analyzer to increase precision by
automatically pruning paths.
}%
% Therefore, we integrate into our analysis \RedSoundRSE a product with such a
% reachability analysis and use it to dynamically prune the paths that
% the relational symbolic execution generates, which not only makes it more
% precise but also more efficient.

\modification{%
This combination of AI and SE is referred to as \RedSoundSE,
defined in \sref{5:sabs}, and is later integrated into the final analysis \RedSoundRSE.
}%

\section{\SoundSE: Sound symbolic execution}
\label{sec:single} \label{s:4:ssym}
We now define a type of symbolic execution, named \SoundSE, as it serves as a basis
for not only \SoundRSE but also \RedSoundSE---the product of \SoundSE with abstract domains.

\paragraph{Symbolic execution states.}
The core principle of symbolic execution is to map program variables into
expressions made of \emph{symbolic values} that denote the initial value
of the program variables.
We let $\SVal = \{ \syvarx, \syvary, \ldots \}$ denote the set of symbolic
values and note for clarity $\syvarx$ the symbolic value associated to
program variable $\varx$ (not to be confused with concrete values).
 A \emph{symbolic store} is a function $\sstore$ from program variables to
\emph{symbolic expressions} the set of which is noted $\SymEx$, namely
expressions defined like the programming language expressions using
symbolic values instead of program variables.
We write $\SMem = \partsof{\Var \rightarrow \SymEx}$ for the set of symbolic
stores and write $[ \varx \leadsto \msingle{\syvarx}, \ldots]$
for an explicitly given symbolic store.
To tie properly symbolic stores and concrete stores, we need to relate
symbolic values and concrete values.
To this end, we let a \emph{valuation} be a function $\valua: \SVal
\longrightarrow \Val$.
Moreover, given a symbolic expression $\symex$, we let $\esem{\symex}$ be
a partial function that maps a valuation $\valua$ to the value obtained
when evaluating the expression obtained by replacing each symbolic
value $\syvarx$ in $e$ with $\valua(\syvarx)$.
We can now express the concretization of symbolic stores:
% we note $\gammasc$ the \emph{symbolic store concretization function}
% that maps a symbolic store into the set of pairs made of a memory and
% a valuation that realize it:
% \[
% \begin{array}{rlcl}
%   \gammasc:
%   & \SMem
%   & \longrightarrow
%   & \partsof{\Mem \times (\SVal \rightarrow \Val)}
%   \\
%   & \sstore
%   & \longmapsto
%   & \{ (\mem,\valua) \mid \forall \varx \in \Var, \;
%   \mem(\varx) = \esem{\sstore(\varx)}(\nu) \}
%   \\
% \end{array}
% \]
\begin{definition}[Symbolic store concretization]
  \label{d:2:gammam}
  The \emph{symbolic store concretization}, $\gammasc:\SMem \longrightarrow \partsof{\Mem \times (\SVal \rightarrow \Val)}$,  maps a symbolic store to the set of pairs made of a store and
  a valuation that realize it, i.e. $\gammasc(\sstore)= \{ (\mem,\valua) \mid \forall \varx \in \Var, \;
    \mem(\varx) = \esem{\sstore(\varx)}(\nu) \}$.
  \iffalse
  \[
  \begin{array}{rlcl}
    \gammasc:
    & \SMem
    & \longrightarrow
    & \partsof{\Mem \times (\SVal \rightarrow \Val)}
    \\
    & \sstore
    & \longmapsto
    & \{ (\mem,\valua) \mid \forall \varx \in \Var, \;
    \mem(\varx) = \esem{\sstore(\varx)}(\nu) \}
    \\
  \end{array}
  \]
  \fi
\end{definition}
To precisely characterize the outcome of an execution path, 
a symbolic store is too abstract. Hence, SE also utilizes a
symbolic expression to constrain the store, referred to as \emph{symbolic path},  that accounts for the conditions encountered during a path.
%execution.
%SE requires not only
%a symbolic store denoting the final values of variables, but also a
% formula that accounts for the conditions encountered during the
%execution.
A \emph{symbolic precise store} is a pair $\sstate = (\sstore,\spath)$
where $\sstore \in \SMem$ and $\spath$ is a symbolic path. 
We write $\Sstate$ for the set of symbolic precise stores.
Their meaning is defined as follows:
% The \emph{symbolic state concretization} $\gammak$ is defined by:
% \[
% \begin{array}{rlcl}
%   \gammak:
%   & \Sstate
%   & \longrightarrow
%   & \partsof{\Mem \times (\SVal \rightarrow \Val)}
%   \\
%   & (\sstore,\spath)
%   & \longmapsto
%   & \{ (\mem,\valua) \in \gammasc(\smap)
%   \mid \esem{\spath}(\valua) = \true \}.
%   \\
% \end{array}
% \]
\begin{definition}[Symbolic precise store concretization]
  \label{d:3:gammak}
   The \emph{symbolic precise store concretization}, $\gammak:\Sstate
    \longrightarrow \partsof{\Mem \times (\SVal \rightarrow \Val)}$, is defined by $
    \gammak(\sstore,\spath) =\{ (\mem,\valua) \in \gammasc(\smap)
    \mid \esem{\spath}(\valua) = \true \} $.

\iffalse  
  The \emph{symbolic state concretization} $\gammak$ is defined by:
  \[
  \begin{array}{rlcl}
    \gammak:
    & \Sstate
    & \longrightarrow
    & \partsof{\Mem \times (\SVal \rightarrow \Val)}
    \\
    & (\sstore,\spath)
    & \longmapsto
    & \{ (\mem,\valua) \in \gammasc(\smap)
    \mid \esem{\spath}(\valua) = \true \}.
    \\
  \end{array}
  \]
  \fi
\end{definition}
\begin{example}[Symbolic precise store]
  \label{e:1:sstate}
  We consider Program~\ref{fp:a}.
  Symbolic execution needs to cover two paths corresponding to each of the
  branches of the condition statement, i.e., depending on the sign of
  $\syvarpriv$.
  Therefore, symbolic execution should produce the precise stores $(\sstore_0,
  \syvarpriv > 0)$ and $(\sstore_1, \syvarpriv \leq 0)$, where $\sstore_0 =
  \sstore_1 = [\vary \leadsto \msingle{5}, \varpriv \leadsto \msingle{\syvarpriv}]$.
\end{example}

\paragraph{Symbolic execution step.}
The main piece of the symbolic execution algorithm is the step relation,
which closely follows the small step semantics of the programs.
We define it by a transition relation $\sse$ between \emph{symbolic
execution states} that are made of a program command and a
symbolic precise store.
Before we write down the analysis $\sse$, we need a few definitions.

First, we define the symbolic evaluation of an expression or condition in
a symbolic store, which produces a symbolic expression.
We note $(\prexpr, \sstore) \seval \symex$ the evaluation of $\prexpr$ into
symbolic expression $\symex$ in symbolic store $\sstore$.
Usually, this evaluation step boils down to the substitution of the
variables in $\prexpr$ with the symbolic expressions they are mapped
to in $\sstore$, possibly with some simplifications.

%% TODO: somewhere have a citation to SMT solvers
Second, we define the conservative satisfiability test of a symbolic path.
This step is usually performed by an external tool such as an SMT solver,
so we do not detail its internals here.
We note that this test may conservatively return as a result that a
symbolic path \emph{may} be satisfiable.
We note $\may{\spath}$ when $\spath$ may be satisfiable.

\begin{figure}[t]
  \[
  \begin{array}{c}
    \inferrule*[leftstyle={\footnotesize \sc},left=s-assign]{
      (\prexpr, \sstore) \seval \symex
    }{
      (\varx := \prexpr, (\sstore, \spath))
      \sse
      (\kwskip, (\sstore[\varx \leadsto \msingle{\symex}], \spath))
    }
    \\[0.35ex]
    \inferrule*[leftstyle={\footnotesize \sc},left=s-if-t]{
      (\prbcond, \sstore) \seval \beta \\
      \spath' \triangleq \spath \wedge \beta \\
      \may{\spath'} \\
    }{
      (\stmtif{\prbcond}{\prcom_0}{\prcom_1},
      (\sstore, \spath))
      \sse
      (\prcom_0, (\sstore, \spath))
    }
    \\[0.35ex]
    \inferrule* [leftstyle={\footnotesize \sc},left=s-loop-t]{
      (\prbcond, \sstore) \seval \beta \\
      \spath' \triangleq \spath \wedge \beta \\
      \may{\spath'} \\
    }{
      (\stmtwhile{\prbcond}{\prcom}, (\sstore, \spath))
      \sse
      (\stmtseq{\prcom}{\stmtwhile{\prbcond}{\prcom}}, (\sstore, \spath))
    }
    \\[0.35ex]
    \inferrule* [leftstyle={\footnotesize \sc},left=s-loop-f]{
      (\prbcond, \sstore) \seval \beta \\
      \spath' \triangleq \spath \logand \neg\beta \\
      \may{\spath'} \\
    }{
      (\stmtwhile{\prbcond}{\prcom}, (\sstore, \spath))
      \sse
      (\kwskip, (\sstore, \spath))
    }
    \\
  \end{array}
  \]
  \caption{Symbolic execution step relation: a few selected rules}
  \label{f:4:ssymstep}
\end{figure}
We now turn to the rules in \fref{4:ssymstep}.
Rule \textsc{s-assign} simply updates the symbolic store with a new
symbolic expression for the assigned variable.
In rule \textsc{s-if-t}, if the guard evaluation $\beta$ is satisfiable, the true branch is accessed
and $\beta$ is added to the symbolic path.
% enters the true branch of a condition statement
% while updating the symbolic path. By symbolically evaluating the guard to $\symex$,  with the symbolic expression derived
% by evaluating the condition when the new symbolic path is satisfiable.
% is not proved
% to be insatisfiable.
Finally, rules \textsc{s-loop-t} and \textsc{s-loop-f} follow similar
principles as rule \textsc{s-if-t} in the case of loops.
% Symbolic execution steps are sound in the following sense:
\modification{We formalize the soundness of execution steps:}
\begin{theorem}[Soundness of a single symbolic execution step]
  \label{t:1:soundone}
  Let $(\prcom, \mem)$ and $(\prcom', \mem') \in \Conf$ be two states such that $(\prcom, \mem) \sem (\prcom', \mem')$, $\sstate
  \in \Sstate$ a symbolic precise store, and $\valua$ be a valuation
  such that $(\mem, \valua) \in \gammak( \sstate )$.
  Then, there exists a symbolic precise store $\sstate'$ such that
  $(\mem', \valua) \in \gammak( \sstate' )$ and $(\prcom, \sstate) \sse
  (\prcom', \sstate')$.
\end{theorem}

\paragraph{Sound depth bounded symbolic execution.}
Clearly, the exhaustive application of the symbolic execution step
relation defined in \fref{4:ssymstep} would not terminate.
Therefore, common symbolic execution tools typically abort the exploration
when they reach some sort of bound on execution lengths.
This result is clearly unsound as longer executions are simply ignored.
Alternatively, it is possible to over-approximate the set of precise stores that
may be reachable when the bound is met.
We formalize this approach here.

Essentially, symbolic states need to be augmented with two
additional pieces of information, namely a boolean so-called
\emph{precision flag} which states whether symbolic execution has
performed any over-approximation due to exhausting the bound, and a
bound control field, called \emph{counter}.
We define set $\Counter$ as the set of counters, with a special element $\counteri \in \Counter$ that denotes the
initial counter status with respect to bound control.
To operate over counters, we require a function $\ctrstep$ which inputs
two commands $\prcom$, $\prcom'$, and a counter $w$.
It produces
a result of the form $(b,w')$ where $b$ is a boolean, and $w'$ is the next counter.
Value $b$ is $\true$ if and only if a step from $\prcom$ to $\prcom'$ can be done without
exhausting the iteration bounds, and with the new counter $w'$.
If $b$ is $\false$, the iteration bound has been reached and the state needs to be
over approximated.

To perform the over approximation, a function $\sstatemodify$ is required. The function inputs a
symbolic store and a command, and returns a new symbolic store $\sstore'$ such that:
\begin{compactitem}
\item \modification{$\sstore'$ maps each program variable that is considered to be ``modified''
  (by a sound over approximation of the set) in $\prcom$ to a \emph{fresh} symbolic value;}
\item $\sstore'$ maps all the other program variables to their image in the original store.
\end{compactitem}
\begin{example}[Loop iteration bounding]
  \label{e:2:loopbnd}
  The most typical way to bound symbolic execution limits the number
  of iteration of each loop to pre-defined number $k$.
  Then, $\Counter$ consists of stacks of integers, $\counteri$ is the
  empty stack, and $\ctrstep$ adds a zero on top of the stack when
  entering a new loop and pops the value on top of the stack when
  exiting a loop.
  More importantly, it increments the value $n$ at the top of the
  stack when $n \leq k$ and moving to the next iteration (rule
  \textsc{s-loop-t});
  on the other hand, when $n > k$, it pops $n$ and returns the $\false$
  precision flag.
\end{example}
To ensure termination, $\Counter$ and $\ctrstep$ should satisfy the following
\emph{well-foundedness} property:
for any infinite sequence of commands $(\prcom_i)_{i}$ the infinite
sequence $(\counter_i)_{i}$ defined by $\ctrstep(\prcom_{i},\prcom_{i+1},
\counter_i) = (\true, \counter_{i+1})$ should be stationary, \modification{which we assume here.}

\begin{figure}[t]
  \[
  \begin{array}{c}
    \inferrule*[leftstyle={\footnotesize \sc},left=s-next]{
      (\prcom, \sstate) \sse (\prcom', \sstate') \\
      \ctrstep(\prcom, \prcom', \counter) = (\true, \counter')
    }{
      (\prcom, \sstate, \counter, b)
      \sse
      (\prcom', \sstate', \counter', b) \\
    }
    \\[0.35ex]
    \inferrule*[leftstyle={\footnotesize \sc},left=s-approx-many]{
      (\prcom, \sstate) \sse (\prcom', \sstate') \\
      \ctrstep(\prcom, \prcom', \counter) = (\false, \counter') \\
      \sstore'' = \sstatemodify( \sstore, \prcom )
    }{
      (\prcom, (\sstore, \spath), \counter, b)
      \sse
      (\kwskip, (\sstore'', \spath), \counter', \false) \\
    }
    \
  \end{array}
  \]
  \caption{\SoundSE: Sound bounded symbolic execution step relation}
  \label{f:5:ssymmax}
\end{figure}
Based on these definitions, \emph{depth bounded symbolic execution} is
defined by a transition relation over 4-tuples made of a command, a
symbolic state, an element of $\Counter$, and a boolean, referred to as symbolic state.
We overload the notation $\sse$ for this relation, which is defined based
on the previously defined $\sse$.
The rules are provided in \fref{5:ssymmax}:
\begin{asparaitem}
\item Rule \textsc{s-next} carries out an atomic step of symbolic
  execution that requires no over approximation; function
  $\ctrstep$ returns the precision
  flag $b$ and a new counter;
\item Rule \textsc{s-approx-many} carries out a global approximation step;
  indeed, as $\ctrstep$ returns $\false$, the function $\sstatemodify$ is
  applied to the symbolic state to over-approximate the effect of an
  arbitrary number of steps of execution of $\prcom$;
  alongside with the new counter state the $\false$ precision is propagated
  forward.
\end{asparaitem}
Under the well-foundedness assumption, exhaustive iteration of the
available symbolic execution rules from any initial symbolic state
will terminate and produce finitely many symbolic states.
To express the soundness of this algorithm, we need to account for the
creation of symbolic values by function $\sstatemodify$, which means
that valuations also need to be extended.
To this end, we note $\valua \valuaprec \valua'$ when the domain of
valuation $\valua$ is included into that of $\valua'$ and when both
$\valua$ and $\valua'$ agree on the intersection of their domains.
We now obtain the following soundness statement:
\begin{theorem}[Soundness of any sequence of single symbolic execution steps]
  \label{t:2:soundmany}
  Let $(\prcom, \mem) \in \Conf$ be a state and $\mem'$ be a store
  such that $(\prcom, \mem) \sems (\kwskip, \mem')$.
  Let $\sstate \in \Sstate$ be a symbolic precise store and $\valua$ be a
  valuation such that $(\mem, \valua) \in \gammak( \sstate )$.
  Let $\counter \in \Counter$ be a counter.
  Then, there exists a symbolic precise store $\sstate'$, a valuation $\valua'$,
  and a counter $\counter' \in \Counter$ such that
  $\valua \valuaprec \valua'$,
  $(\mem', \valua') \in \gammak( \sstate' )$, and
  $(\prcom, \sstate, \counter, b) \sses (\kwskip, \sstate', \counter', b')$.
\end{theorem}
The proof of this theorem follow from \tref{1:soundone} (steps where
$\ctrstep$ returns $\true$), and a global induction on the
command $\prcom$ when rule \textsc{s-approx-many} applies.
\begin{example}[Symbolic execution]
  \label{e:3:ssym}
  For program \ref{fp:a}, symbolic execution returns the symbolic stores 
  shown in \eref{1:sstate}.
  We assume the bounding of \eref{2:loopbnd} and consider program \ref{fp:b}.
  Then, symbolic execution generates the symbolic store $[ \varz \leadsto
  \msingle{\syvarz}, \vari \leadsto \msingle{\syvari'}, \varpriv \leadsto \msingle{\syvarpriv'}]$
  with precision flag $\false$, and
  where $\syvari'$, $\syvarpriv'$ are fresh symbolic values generated by rule
  \textsc{s-approx-many}.
\end{example}

\paragraph{Refutation up to a bound.}
A very desirable feature of symbolic execution is the ability to produce
counter-examples up to a bound.
This feature stems from a bounded refutation result, which states
that, when symbolic execution produces a final state for which
the final precision flag is $\true$, and such that the symbolic path is
satisfiable, then a matching concrete execution can be found.
\modification{From the final state, the SMT solver can compute a refutation
model.}
% \begin{theorem}[Refutation up to a bound]
%   \label{t:3:refute}
%   Let $\prcom$ be a command, $\sstate, \sstate' \in \Sstate$ be two
%   precise stores, and $\counter, \counter' \in \Counter$.
%   Let $(\mem, \valua) \in \gammak( \sstate )$.
%   If $(\prcom, \sstate, \counter, \true) \sses (\kwskip, \sstate',
%   \counter', \true)$, then there exists a unique store $\mem' \in \Mem$
%   such that
%   $(\mem', \valua) \in \gammak( \sstate' )$ and
%   $(\prcom, \mem) \sems (\kwskip, \mem')$.
% \end{theorem}
\begin{theorem}[Refutation up to a bound]
  \label{t:3:refute}
  Let $\prcom$ be a command, $\sstate, \sstate' \in \Sstate$ be two
  precise stores, $\counter, \counter' \in \Counter$, such that
  $(\prcom, \sstate, \counter, \true) \sses (\stmtskip, \sstate',
  \counter', \true)$.
  Then, for all $(\mem', \valua') \in \gammak(\sstate')$, it exists $(\mem, \valua) \in \gammak(\sstate)$ such that
  $(\prcom, \mem) \sems (\stmtskip, \mem')$.
\end{theorem}
This result follows from the fact that rule \textsc{s-approx-many} is never
applied in the symbolic execution and from an induction on the sequence of
\textsc{s-next} steps.
\begin{example}[Symbolic execution completeness up to a bound]
  \label{e:4:csym}
  We consider the cases discussed in \eref{3:ssym}.
  Using the bounding of \eref{2:loopbnd}, the result produced for program
  \ref{fp:a} is complete whereas that for program \ref{fp:b} generates
  some final symbolic state with precision flag $\false$, hence
  for which \tref{3:refute} does not apply.
\end{example}
%% TODO: add that symbolic execution generates finitely many states

\section{\modification{\RedSoundSE: Sound SE combined with abstract states}}
\label{s:5:sabs}
We now extend \SoundSE with the ability to
use the properties inferred by abstract interpretation.
This combined symbolic execution is referred to as \RedSoundSE, making reference to
the reduced product between \SoundSE and an AI based analysis.

\paragraph{\modification{Abstraction of store and static analysis.}}
In the following, we assume that an \emph{abstract domain}~\cite{cc:popl:77}
$\Astate$ describing sets of stores is fixed, together with a
concretization function $\gammast: \Astate \longrightarrow \partsof{\Mem}$.
We assume the existence of an element $\bot \in \Astate$ such that
$\gammast( \bot ) = \emptyset$.
Additionally, we require the two following sound abstract post-condition
functions for basic operations. Function $\esem{\ }$ will be overloaded to
replace any variable $\varx$ for its mapped value in a store $\mem(\varx)$:
\begin{asparaitem}
\item \emph{abstract assignment} $\absassign_{\varx,\prexpr}: \Astate
  \longrightarrow \Astate$ is parameterized by a variable $\varx$ and an
  expression $\prexpr$ and is such that %, for all abstract state $\astate
  $\forall \astate
  \in \Astate$, $\{ \mem[\varx \mapsto \esem{\prexpr}(\mem)] \mid \mem \in
  \gammast( \astate ) \} \subseteq \gammast( \absassign_{\varx,\prexpr}(
  \astate ) )$.
\item \emph{abstract condition} $\absguard_{\prbcond}: \Astate \longrightarrow
  \Astate$ is parameterized by a boolean expression $\prbcond$ and is such
  that %, for all abstract state $\astate \in \Astate$, $\{ \mem \in \gammast(
  $\forall \astate \in \Astate$, $\{ \mem \in \gammast(
  \astate ) \mid \esem{\prbcond}( \mem ) = \true \} \subseteq \gammast(
  \absguard_{\prbcond}( \astate ) )$.
\end{asparaitem}
Based on these operations, the definition of a \emph{sound abstract execution
  step} relation $\ssa$ is straightforward.
We show two rules in \fref{6a:astep}.
The rules match those of $\sem$ (\sref{3:lang}) and are sound with respect
to it.
In the following, $\Astate$ is assumed to be a parameter of the analysis.
It may consist of any numerical abstraction, such as the
interval abstract domain~\cite{cc:popl:77} or the domain of convex
polyhedra~\cite{ch:popl:78}.
Moreover, the application of standard widening technique~\cite{cc:popl:77}
allows to define a \emph{static analysis} function $\asema{\prcom}:
\Astate \longrightarrow \Astate$ that is sound in the sense that, for
all command $\prcom$ and all abstract state $\astate$,
% \[
$
\{ \mem' \in \Mem \mid \exists \mem \in \gammast(\astate), \;
(\prcom, \mem) \sses (\kwskip,\mem') \}
\subseteq \gammast( \asema{\prcom}( \astate ) )
$
% \]
\begin{figure}[t]
  \begin{center}
    \subfigure[Abstract execution step selected rules]{
      \label{f:6a:astep}
      \(
      \begin{array}{c}
        \inferrule*[leftstyle={\footnotesize \sc},left=a-assign]{
          \astate' \triangleq \absassign_{\varx,\prexpr}( \astate )
        }{
          (\varx := \prexpr, \astate)
          \ssa
          (\kwskip, \astate')
        }
        \qquad
        \inferrule*[leftstyle={\footnotesize \sc},left=a-if-t]{
          \astate' \triangleq \absguard_{\prbcond}( \astate ) \\
          \astate' \not= \bot
        }{
          (\kwif\ \prbcond\ \kwthen\ \prcom_0\ \kwelse\ \prcom_1,
          \astate)
          \ssa
          (\prcom_0, \astate')
        }
        \\
      \vspace{0.01em}
      \end{array}
      \)
    } \\
    \subfigure[Product of symbolic execution and static analysis]{
      \label{f:6b:rpstep}
      \(
      \begin{array}{c}
        \inferrule*[leftstyle={\footnotesize \sc},left=s-a-next]{
          (\prcom, \sstate, \counter, b)
          \sse
          (\prcom', \sstate', \counter', b) \\
          \ctrstep(\prcom, \prcom', \counter) = (\true, \counter') \\
          (\prcom, \astate) \ssa (\prcom', \astate') \\
          (\sstate'', \astate'') \triangleq \reduction( \sstate', \astate' )
        }{
          (\prcom, \sstate, \astate, \counter, b)
          \rsse
          (\prcom', \sstate'', \astate'', \counter', b) \\
        }
        \\[2ex]
        \inferrule*[leftstyle={\footnotesize \sc},left=s-a-approx-many]{
          (\prcom, \sstate) \sse (\prcom', \sstate') \\
          \ctrstep(\prcom, \prcom', \counter) = (\false, \counter') \\
          \sstate'' = \sstatemodify( \sstate, \prcom ) \\
          \astate' = \asema{\prcom}( \astate ) \\
          (\sstate''', \astate''') \triangleq \reduction( \sstate'', \astate')
        }{
          (\prcom, (\sstate, \astate), \counter, b)
          \rsse
          (\kwskip, (\sstate''', \astate'''), \counter', \false) \\
        }
        \\
      \vspace{0.01em}
      \end{array}
      \)
    }
  \end{center}
    \vspace{-1.5em}
  \caption{Abstract execution step and product with symbolic execution}
  \label{f:6:asymstep}
\end{figure}

\paragraph{\modification{Reduced product of symbolic precise stores and abstract states.}}
Reduced product~\cite{cc:popl:79} aims at expressing precisely conjunctions
of constraints expressed in distinct abstract domains.
\modification{We let a precise product store be a pair $(\sstate,\astate)
  \in \Sstate \times \Astate$.}
In our case, the definition needs to be adapted slightly as symbolic
execution and abstract domain $\Astate$ do not abstract exactly the same
objects:
% The \emph{product abstract domain} consists of the set $\Sstate \times
% \Astate$ and the concretization function $\gammasxa: \Sstate \times
% \Astate \longrightarrow \partsof{\Mem \times (\SVal \rightarrow \Val)}$
% defined as follows:
% \[
% \gammasxa:
% ( \sstate, \astate ) \longmapsto
% \{ (\mem, \valua) \in \gammasc( \sstate )
% \mid \mem \in \gammast( \astate ) \}
% \]
\begin{definition}[\modification{Product domain}]
  \label{d:4:ssymabs}
  The \emph{product abstract domain} consists of the set $\Sstate \times
  \Astate$ and the concretization function $\gammasxa: \Sstate \times
  \Astate \longrightarrow \partsof{\Mem \times (\SVal \rightarrow \Val)}$
  defined as follows:
  %\[
  $
  \gammasxa:
  ( \sstate, \astate ) \longmapsto
  \{ (\mem, \valua) \in \gammasc( \sstate )
  \mid \mem \in \gammast( \astate ) \}
  $
  % \]
\end{definition}
\modification{%
In a precise product store $( \sstate, \astate )$, the goal is to enhance precision by exchanging
information between $\sstate$ and $\astate$.
This is done through a {\bf reduction} function, which rewrites an abstract element with another of
equal concretization, but that supports more precise analysis operations.
This implies that $(\gammasxa \circ \reduction) ( \sstate, \astate ) = \gammasxa ( \sstate, \astate )$.
}%
This requires the abstract domain $\Astate$ to
support a function $\absconstr$ that maps an abstract state $\astate$ to
a logical formula over program variables and entailed by $\astate$, namely
such that, if $\mem \in \gammast(\astate)$ then $\mem$ satisfies formula
$\absconstr(\astate)$.
Some abstract domains\modification{---specifically intervals and abstract polyhedra---}utilize an internal representation based on
conjunction of constraints, in which case $\absconstr$ is trivial.
Then, $\reduction: \Sstate \times \Astate \longrightarrow \Sstate \times
\Astate$ is defined by:
\[
\reduction((\sstore,\spath),\astate) \triangleq ((\sstore,\spath'),\astate)
\quad \text{where} \quad
\spath' \triangleq \spath \logand \absconstr(\astate)[\vec{\varx} \mapsto
\sstore(\vec{\varx})]
\]
Note that $[\vec{\varx} \mapsto \sstore(\vec{\varx})]$ in the above definition, symbolizes the
replacement of each program variable present
in $\absconstr(\astate)$ into its definition in $\sstore$; this step follows
from the fact that $\astate$ constrains program variables whereas $\spath$
constrains valuations.
This general reduction function may be refined into a more precise one,
where the resulting symbolic path is simplified, possibly to the $\false$
formula.
Furthermore, this reduction only modifies the symbolic path $\spath$, but it
is possible to define a reduction operation that also rewrites the abstract state $\astate$.

\paragraph{Reduced product symbolic execution.}
The product analysis, namely \RedSoundSE, takes the form of an extension of the symbolic
execution function of \fref{5:ssymmax}.
\modification{%
The new states are still 4-tuples, but the symbolic precise
store component $\sstate$ is now replaced with a precise product store $(\sstate,\astate)$.
}%
The transition relation $\ssasxa$ between such states consists of
two rules that are shown in \fref{6b:rpstep} and that extend those in
\fref{5:ssymmax}.
% \erasable{
% Rule \textsc{s-a-next} corresponds to a regular symbolic execution step,
% and in this case, a similar abstract execution step is performed.
% }
\modification{%
In rule \textsc{s-a-many} (applied when exploration bound is met) aside from $\sstatemodify$, 
the loop is calculated over the abstract state and then the reduction function is applied.
}%

In both cases, the sound $\reduction$ operator may be applied.
In practice, for the sake of efficiency, it can be computed and applied in
a lazy manner that is, only for specific steps (typically \textsc{s-a-many}
and for branching commands).
% \begin{example}[Product analysis]
%   In program \ref{fp:c}, rule \textsc{s-a-many} allows to maintain the
%   information that, on all paths, the symbolic expression $\sstore(\varpriv)$
%   holds a positive value, hence only the true branch of the condition may
%   be visited.
% \end{example}
\modification{%
\begin{example}[Product analysis]
For program \ref{f:3:ex}, assuming $\vari < 10$, and then when exiting the loop, an intervals abstract state will hold
two constraints $\astate = \{ \vari = 10 ;\ \varpriv \geq 2 \}$.
Assuming a symbolic precise store $\sstate = (\sstore, \spath)$ with
$\sstore = [ \vari \rightarrow \syvari ;\  \varpriv \rightarrow \syvarpriv ]$, the
abstract constraints can be fitted to a symbolic path $\spath'$ as follows:
$ \spath' \triangleq\ \spath\ \wedge\ \syvari = 10\ \wedge\ \syvarpriv \geq 2 $.
A more detailed execution trace is given in Appendix \ref{app:example}.
\end{example}
}%

\paragraph{Soundness and refutation property.}
The \RedSoundSE analysis defined in the previous paragraph satisfies the same
soundness (\tref{2:soundmany}) and refutation (\tref{3:refute}) properties
as standard symbolic execution, so we do not give the theorems again.
% \erasable{for
% the sake of space and as the statements are very similar.}

\section{\SoundRSE: Sound relational symbolic execution}
\label{section:double} \label{s:6:dsym}
As discussed in \sref{2:over}, security properties like noninterference
require to reason over \emph{pairs} of execution traces thus we now set
up a \emph{sound relational symbolic execution} technique that constructs
pairs of executions.
This analysis will be regarded as \SoundRSE. 

\paragraph{Assumption.}
To keep notations lighter, we assume in this section and the next that
the bounding counter step function $\ctrstep$ only affects loops, namely
$\ctrstep(\prcom,\prcom',\counter) = \counter$ whenever
$\prcom$ is not a loop command.
Moreover, we do not include the product with the numerical abstract
state (as in \sref{5:sabs}) in the following definitions.
Since it can be added in a seamless manner, we omit it here to keep
formal statements lighter.

%% TODO: should we name this "symbolic double states" ?
\paragraph{\modification{Precise relational stores.}}
\modification{We first define the notions of relational expression, relational store, and
precise relational store.}
\begin{definition}[Relational and precise relational stores]
  \label{d:5:dsym}
  A \emph{relational symbolic expression} is an element defined by the
  grammar: $\rsymex ::= \msingle{\symex} | \mpair{\symex}{\symex}$
  where $\symex$ ranges over the set $\SymEx$ of symbolic expressions.
  We write $\DSymEx$ for the set of relational symbolic expressions.
  A \emph{relational symbolic store} $\rsstore$ is a function from
  variables to relational symbolic expressions.
  We let $\DSMem = \Var \rightarrow \DSymEx$ stand for their set.
  \modification{%
  Finally, a \emph{precise relational store} $\rsstate$ is a pair
  $(\rsstore, \spath) \in \DSState$.
  }%
\end{definition}
%
% Before we define the concretizations of $\DSMem$ and $\DSState$, we need
% to introduce two operations over $\DSMem$:
\modification{Before we define concretizations of $\DSMem$ and $\DSState$, we need to introduce
two operations:}
\begin{asparaitem}
\item The \emph{projections} $\projl, \projr$ map relational symbolic stores
  into symbolic stores.
  They are defined in a pointwise manner, as follows:
  if $\rsstore(\varx) = \msingle{\symex}$ then $\projl(\rsstore)(\varx) = \projr(
  \rsstore)(\varx) = \symex$ and if $\rsstore(\varx) = \mpair{\symex_0}{\symex_1}$, then
  $\projl(\rsstore)(\varx) = \symex_0$ and
  $\projr(\rsstore)(\varx) = \symex_1$.
  We overload the $\projl,\projr$ notation and also apply it to double
  symbolic expressions:
  $\projl( \msingle{\symex} )= \projr( \msingle{\symex} ) =
  \symex$ and if $\rsymex = \mpair{\symex_0}{\symex_1}$, then $\projl(\rsymex) = \symex_0$ and $\projr(\rsymex)
  = \symex_1$.
\item The \emph{pairing} $\pairing{\sstore_0}{\sstore_1}$ of two symbolic
  stores $\sstore_0$ and $\sstore_1$ is a relational symbolic store defined
  such that, for all variable $\varx$,
  \[
  \pairing{\sstore_0}{\sstore_1}( \varx ) =
  \left\{
    \begin{array}{lll}
      \msingle{\symex}
      & \quad & \text{if } \sstore_0( \varx ) \text{ and } \sstore_1( \varx )
      \text{ are provably equal to } \symex \in \SymEx
      \\
      \mpair{\sstore_0(\varx)}{\sstore_1(\varx)}
      & \quad & \text{otherwise}
      \\
    \end{array}
  \right.
  \]
  where the notion of ``provably equal'' may boil down to syntactic
  equality of symbolic expressions or involve an external proving tool.
\end{asparaitem}
We can now define the concretization functions:
% The \emph{concretization of relational symbolic stores} $\dgammam$
% and \emph{concretization of relational symbolic states} $\dgammak$ are
% defined by:
% \[
% \begin{array}{rlcl}
%   \dgammam:
%   & \DSMem
%   & \longrightarrow
%   & \partsof{\Mem \times \Mem \times (\SVal \rightarrow \Val)}
%   \\
%   & \rsstore
%   & \longmapsto
%   & \{ (\mem_0,\mem_1,\valua) \mid \forall \varx \in \Var, \;
%   \forall i \in \{0,1\}, \;
%   \mem_i(\varx) = \esem{\proji{i}(\rsstore)(\varx)}(\nu) \}
%   \\[0.5ex]
%   \dgammak:
%   & \DSState
%   & \longrightarrow
%   & \partsof{\Mem \times \Mem \times (\SVal \rightarrow \Val)}
%   \\
%   & (\rsstore,\spath)
%   & \longmapsto
%   & \{ (\mem_0,\mem_1,\valua) \in \dgammam(\rsstore)
%   \mid \esem{\spath}(\valua) = \true \}.
%   \\
% \end{array}
% \]
\begin{definition}[Concretization functions]
  \label{d:6:dgamma}
  The \emph{concretization of relational stores} $\dgammam$
  and \emph{concretization of precise relational stores} $\dgammak$ are
  defined by:
  \[
  \begin{array}{rlcl}
    \dgammam:
    & \DSMem
    & \longrightarrow
    & \partsof{\Mem \times \Mem \times (\SVal \rightarrow \Val)}
    \\
    & \rsstore
    & \longmapsto
    & \{ (\mem_0,\mem_1,\valua) \mid \forall \varx \in \Var, \;
    \forall i \in \{0,1\}, \;
    \mem_i(\varx) = \esem{\proji{i}(\rsstore)(\varx)}(\nu) \}
    \\[0.5ex]
    \dgammak:
    & \DSState
    & \longrightarrow
    & \partsof{\Mem \times \Mem \times (\SVal \rightarrow \Val)}
    \\
    & (\rsstore,\spath)
    & \longmapsto
    & \{ (\mem_0,\mem_1,\valua) \in \dgammam(\rsstore)
    \mid \esem{\spath}(\valua) = \true \}.
    \\
  \end{array}
  \]
\end{definition}
\begin{example}
  We consider program~\ref{fp:a} (\SoundSE was
  discussed in \eref{1:sstate}).
  To cover pairs of executions that start with the same value for low
  variable $\vary$ but possibly distinct values for high variable
  $\varpriv$, relational symbolic execution should cover four pairs of paths.
  These four paths have the same relational symbolic store $[\varpriv
  \leadsto \mpair{\syvarpriv_0}{\syvarpriv_1}, \vary \leadsto \msingle{5}]$ and differ only in the
  symbolic path components.
  For instance, when the first execution takes the true branch of the
  condition and the second the false branch, the symbolic path is
  $\syvarpriv_0 > 0 \logand \syvarpriv_1 \leq 0$.
\end{example}

\paragraph{Relational symbolic execution algorithm.}
Since \SoundRSE aims at describing pairs of executions, it should account for
the case where the two executions follow different control flow paths.
Thus, a relational symbolic state may consist of a single command
when both executions follow the same path, or two commands when they
diverge.
We respectively note these two kinds of states $(\prcom, \rsstate,
\counter, b)$ and $(\stmtseq{(\prcom_0 \bowtie \prcom_1)}{\prcom_2}, \rsstate, \counter, b)$;
in the latter, $\prcom_0$ (resp., $\prcom_1$) denotes the control state
of the first (resp., second) execution, \modification{which they later meet in $\prcom_2$.}
The components $\counter$ and $b$ have the same meaning as in \sref{4:ssym}.
Initial states are of the former sort.

%% TODO: the two if rules should be expanded without using sse
We write $\dse$ for the relational symbolic execution step relation.
A representative selection of the rules are shown in \fref{7:dsymstep}.
% \erasable{
% Rules \textsc{\SRassign} and \textsc{\SRiftt} simultaneously make the
% same step in the two executions that are described.
% }
% The use the $\sreval$ evaluation of program expressions over relational
% symbolic states, that returns a relational symbolic expression.
% Rule \textsc{\SRiftf} describes a point where the two executions being
% constructed diverge, as the left (resp., right) execution follows the
% \true\ (resp., \false) branch of the condition command.
% Divergence is materialized by the appearance of the $\prcom_0 \bowtie
% \prcom_1$ control state.
% Rule \textsc{\SRexit} marks the end of a relational symbolic execution
% sequence, as both executions finish.
% \modification{
% Rules {\sc \SRcompl} and {\sc \SRcompr} handle states where the programs have diverged.
% These two executions meet after executing all the steps of the diverged paths.
% }
Rule \textsc{\SRloopmany} describes a case where approximation is
performed so as to ensure termination and uses the straightforward
extension of $\sstatemodify$ to relational symbolic states.
\newcommand{\dbeta}{\tilde{\beta}}
\begin{figure}[t]
  \[
  \begin{array}{c}
    \inferrule*[leftstyle={\sc},left=\SRexit]{
    }{
      (\kwskip \bowtie \kwskip, (\rsstore, \spath), \counter, b)
      \dse
      (\kwskip, (\rsstore, \spath), \counter, b)
    }
    % \\
    \\[0.75ex]
    \inferrule*[leftstyle={\sc},left=\SRcompr]{
      (\prcom_1, (\projr(\rsstore), \spath), \counter, b)
      \sse
      (\prcom'_1, (\sstore'_1, \spath'), \counter', b')
    }{
      (\kwskip \bowtie \prcom_1, (\rsstore, \spath), \counter, b)
      \dse
      (\kwskip \bowtie \prcom'_1, (\pairing{\projl(\rsstore)}{\sstore'_1},
      \spath'), \counter', b')
    }
    \\[0.75ex]
    \inferrule*[leftstyle={\sc},left=\SRcompl]{
      (\prcom_0, (\projl(\rsstore), \spath), \counter, b)
      \sse
      (\prcom'_0, (\sstore'_0, \spath'), \counter', b')
    }{
      (\prcom_0 \bowtie \prcom_1, (\rsstore, \spath), \counter, b)
      \dse
      (\prcom'_0 \bowtie \prcom_1, (\pairing{\sstore'_0}{\projr(\rsstore)},
      \spath'), \counter', b')
    }
    \\[0.75ex]
    \inferrule*[leftstyle={\sc},left=\SRiftf]{
      (\prbcond, \rsstore) \sreval \dbeta \\
      \spath' = \spath \logand \projl(\dbeta) \logand \neg \projr(\dbeta) \\
      \may{\spath'} \\
    }{
      (\stmtif{\prbcond}{\prcom_0}{\prcom_1},
      (\rsstore, \spath), \counter, b)
      \dse
      (\prcom_0 \bowtie \prcom_1,
      (\rsstore, \spath'),
      \counter, b)
    }
    \\[0.75ex]
    \inferrule*[leftstyle={\sc},left=\SRloopmany]{
      \ctrstep(\stmtwhile{\prbcond}{\prcom},
      (\stmtseq{\prcom}{\stmtwhile{\prbcond}{\prcom}}), \counter) =
      (\false, \counter') \\
      \rsstore'' = \sstatemodify( \rsstore, \prcom ) \\
      (\prbcond, \rsstore'') \sreval \langle \beta_0, \beta_1 \rangle \\
      \spath' \triangleq \spath \logand \neg\beta_0 \logand \neg\beta_1
    }{
      (\stmtwhile{\prbcond}{\prcom},
      (\rsstore, \spath), \counter, b)
      \dse
      (\kwskip, (\rsstore'', \spath'), \counter', \false)
    }
  \end{array}
  \]
  \caption{\SoundRSE: a few selected rules of the relational symbolic execution step relation.}
  \label{f:7:dsymstep}
\end{figure}
%% TODO: should we make a paragrah on how DSE works when using also
%% a numerical abstract domain

\paragraph{Soundness and refutation property.}
\SoundRSE inherits similar soundness and refutation
properties as \SoundSE, as shown in the following
theorems.
\begin{theorem}[Soundness]
  \label{t:4:sounddsym}
  Let $\rsstate \in \DSState$, $\counter \in \Counter$,
  and $b \in \sBool$.
  We let $(\mem_0, \mem_1, \valua) \in \dgammak( \rsstate )$ and
  assume that stores $\mem'_0, \mem'_1$ are such that $(\prcom,
  \mem_0) \sems (\kwskip, \mem'_0)$ and $(\prcom, \mem_1) \sems (\kwskip,
  \mem'_1)$.
  Then, there exists $\rsstate' \in \DSState$, a valuation
  $\valua'$, and a counter state $\counter' \in \Counter$ such that
  $\valua \valuaprec \valua'$, 
  $(\mem'_0, \mem'_1, \valua') \in \gammak( \rsstate' )$, and
  $(\prcom, \rsstate, \counter, b) \dses (\kwskip, \rsstate', \counter', b')$.
\end{theorem}
% \begin{theorem}[Refutation up to a bound]
%   \label{t:5:refute}
%   Let $\prcom$ be a command, $\rsstate, \rsstate' \in \DSState$, $\counter, \counter' \in \Counter$,
%   and $(\mem_0, \mem_1, \valua) \in \dgammak( \rsstate )$.
%   If $(\prcom, \rsstate, \counter, \true) \dses (\kwskip, \rsstate',
%   \counter', \true)$, then there exists stores $\mem'_0, \mem'_1
%   \in \Mem$ such that
%   $(\mem'_0, \mem'_1, \valua) \in \dgammak( \rsstate' )$
%   and
%   $\forall i \in \{ 0, 1 \}, \;
%   (\prcom, \mem_i) \sems (\kwskip, \mem'_i)$.
% \end{theorem}
\begin{theorem}[Refutation up to a bound]
  \label{t:5:refute}
  Let $\prcom$ be a command, $\rsstate, \rsstate' \in \DSState$ be two
  precise stores, $\counter, \counter' \in \Counter$, such that
  $(\prcom, \sstate, \counter, \true) \dses (\stmtskip, \sstate',
  \counter', \true)$.
  Then, for all $(\mem_0', \mem_1', \valua') \in \dgammak(\sstate')$, it exists $(\mem_0, \mem_1, \valua) \in \dgammak(\sstate)$ such that
  $(\prcom, \mem_0) \sems (\stmtskip, \mem_0')$ and $(\prcom, \mem_1) \sems (\stmtskip, \mem_1')$.
\end{theorem}

\paragraph{\SoundRSE-based analysis and noninterference.}
We now assume a program $(\prcom, L)$,
% with body $\prcom$, a set of initially low variables $L$
% and initially high variables $H$,
and show the application of \SoundRSE
analysis to attempt proving noninterference.
The analysis proceeds according to the following steps:
\newcommand{\dseoutput}{\mathcal{O}}
\begin{compactenum}
\item Construction of the initial store $\rsstore_0$ such that,
  for all variables $\varx$ present in $\prcom$, $\rsstore_0(\varx) = \msingle{\syvarx}$
  (\resp, $\rsstore_0(\varx) = \mpair{\syvarx_0}{\syvarx_1}$)
  if $\varx \in L$ (\resp, $\varx \not\in L$), and where $\syvarx$ is
  a fresh symbolic value (\resp, $\syvarx_0, \syvarx_1$ are fresh symbolic
  values).
\item \modification{%
  Exhaustive application of semantic rules from initial state
  $(\prcom, (\rsstore_0, \true), \counteri, \true)$;
  }%
  we let $\dseoutput$ stand for the set of final precise relational stores with their precision
  flags:
  % \[
  $
  \dseoutput \triangleq
  \{ (\rsstate, b) \mid \exists \counter \in \Counter, \;
  (\prcom, (\rsstore_0, \true), \counteri, \true)
  \dse
  (\kwskip, \rsstate, \counter, b) \}.
  $
  % \]
\item \emph{Attempt to prove noninterference} for each symbolic path in
  $\dseoutput$ using an external tool, such as an SMT solver;
  more precisely, given $((\rsstore, \spath), b) \in \dseoutput$,
  \begin{compactitem}
  \item if $\spath$ is not satisfiable, the path is infeasible and can be
    ignored;
  \item if it can be proved that for all variables $\varx \in L$, there is a unique value, i.e.,
  $\projl(\rsstore)(\varx) = \projr(\rsstore)(\varx)$, then the program is noninterferent;
  % low variables share the same value for
  % \item if, for all low variables $\varx \in L$, the constraint
  %   $\projl(\rsstore)(\varx) = \projr(\rsstore)(\varx)$ can be proved
  %   then noninterference is proved for this path;
  % \item poronga
  % \[
  %  \biggl( \bigwedge_{\varx \in L} \projl(\rsstore)(\varx) = \projr(\rsstore)(\varx) \biggr) \wedge \spath
  % \]
  \item if a valuation $\valua$ can be found, such that $\esem{\spath}(
    \valua) = \true$ (the path is satisfiable), and there exists a variable
    $\varx \in L$ such that $\esem{\projl(\rsstore)(\varx)}(\valua) \not=
    \esem{\projr(\rsstore)(\varx)}(\valua)$, and $b = \true$, then
    $\valua$ provides a counter-example refuting noninterference;
  \item finally, if $b = \false$ and neither of the above cases occurs,
    no conclusive answer can be given for this path.
  \end{compactitem}
\end{compactenum}
To summarize, the analyser either proves noninterference (when all paths
are either not satisfiable or noninterferent), or it provides a valuation
that refutes noninterference (when such a valuation can be found for at least
one path), or it does not conclude.
When a refutation is found, this refutation actually defines a real attack.
\begin{example}[Noninterference]
  In the case of program \ref{fp:a}, all paths are low-equal.
  The analysis of program \ref{fp:c} computes at least one interferent
  path if the unrolling bound is set to any strictly positive integer;
  in that case, a model such as the one presented in \sref{2:over} can
  be synthesized by even basic SMT solvers.
  Finally, the program of \fref{3:ex} can be proved noninterferent with
  relational symbolic execution combined with a reduced product with a
  value abstract domain such as intervals (\sref{5:sabs}).
\end{example}

% \vspace{-1.9em}
\modification{%
\section{\RedSoundRSE: Product of \SoundRSE with
  Dependence AI}
}%
% \section{\RedSoundRSE: Reduced product of \SoundRSE with
%   abstract interpretation based static analysis}
\label{s:7:sdep}
As observed in \sref{2:over} some programs like that of \fref{fp:b} can
be analyzed more precisely using conventional dependence analysis than
by bounded symbolic execution (\sref{4:ssym}).
In this section, we set up a novel form of product of abstractions, so
as to benefit from this increase in precision.
This notion of product is generic and does not require to fix a specific
dependency abstraction.
% The final analysis presented in this section is referred to as \RedSoundRSE.
We refer to the final analysis presented in this section as \RedSoundRSE.

\paragraph{Dependence abstraction and static analysis.}
Although dependence abstractions may take many forms, they all characterize
information flows that can be observed by comparing pairs of executions.
For instance, \cite{anst:popl:17} uses a lattice of security levels and
abstract elements map each level to a set of variables. These variables are left
unmodified when the input value of variables of higher levels change.
Other works use relational abstract domains, where relational means that
relations are maintained \emph{across pairs of executions}.
Therefore, we can characterize such analyses with an abstraction of pairs
of stores:
% A \emph{dependence abstraction} is defined by an abstract lattice
% $\Dep$ and a concretization function $\gammadep: \Dep \rightarrow
% \Mem \times \Mem$.
% Moreover, a \emph{sound dependency analysis} is defined by a function
% $\asemd{\prcom}: \Dep \rightarrow \Dep$ such that, for all $\dep \in
% \Dep$, and for all $(\mem_0, \mem_1) \in \gammadep( \dep )$,
% \[
% \{ (\mem'_0, \mem'_1) \in \Mem \times \Mem \mid
% \forall i \in \{ 0, 1 \}, \; (\prcom, \mem_i) \sem (\kwskip, \mem'_i) \}
% \subseteq \gammadep \circ \asemd{\prcom}( \dep ).
% \]
\begin{definition}[Dependence abstraction and analysis]
  \label{d:7:adep}
  A \emph{dependence abstraction} is defined by an abstract lattice
  $\Dep$ from security levels to variables and a concretization function
  \[
  \begin{array}{rlcl}
    \gammadep:
    & \Dep
    & \longrightarrow
    & \partsof{\Mem \times \Mem}
    \\
    & \dep
    & \longmapsto
    & \{ (\mem_0, \mem_1) \in \Mem \times \Mem \mid
    \mem_0 =_{\dep( \Low )} \mem_1 \}
    \\
  \end{array}
  \]
  A \emph{sound dependency analysis} is defined by a function
  $\asemd{\prcom}: \Dep \rightarrow \Dep$ such that, for all $\dep \in
  \Dep$, $(\mem_0, \mem_1) \in \gammadep( \dep )$,
  % \[
  $
  \{ (\mem'_0, \mem'_1) \in \Mem \times \Mem \mid
  \forall i \in \{ 0, 1 \}, \; (\prcom, \mem_i) \sem (\kwskip, \mem'_i) \}
  \subseteq \gammadep \circ \asemd{\prcom}( \dep ).
  $
  % \]
\end{definition}
\begin{example}[Standard dependence based abstraction~\cite{anst:popl:17}]
  \label{e:8:dep}
  The abstraction of \cite{anst:popl:17} is an instance of \dref{7:adep}.
  Let $\{ \Low, \High \}$ be the set of security levels.
  Assume an initial abstract state $\dep$ that captures pairs of concrete stores that are
  low equal for some program $(\prcom, L)$.
  By applying the dependence analysis, if the final dependence state has a low dependency for each
  initially low variable, the program is noninterferent.
  
  In practice such information is computed by forward abstract
  interpretation, using syntactic dependencies for expressions and
  conditions, and conservatively assuming conditions may generate
  (implicit) flows to any operation that they guard.
\end{example}
We note that \dref{7:adep} accounts not only for dependence abstractions
such as that of~\cite{anst:popl:17}.
In particular, \cite{dm:sas:19} proposes a semantic patch analysis which
can also be applied to security properties by using a relational abstract
domain to relate pairs of executions; such analyses use an abstraction that
also writes as in \dref{7:adep}.
In the following, we assume a sound dependence analysis is fixed.

\paragraph{Product of symbolic execution and dependence analysis.}
We now combine dependence analysis and symbolic execution.
For most statements, \SoundRSE rules defined in
\fref{7:dsymstep} introduce no imprecision.
The notable exception is the case where the execution bound is reached
as in rule \textsc{\SRapproxmany}.
Therefore, the principle of the combined analysis is to replace this
imprecise rule with another that uses dependence analysis results to
strengthen relational stores.
First, we introduce two operations to transport information in a sound
manner into and from the dependence abstract domain:
% The \emph{translation from symbolic to dependence} is a function
% $\tausd: \DSMem \rightarrow \Dep$ that is sound in the following
% sense:
% \[
% \forall \rsstore \in \DSMem, \;
% \forall (\mem_0, \mem_1, \valua) \in \dgammam( \rsstore ), \;
% (\mem_0, \mem_1) \in \gammadep \circ \tausd( \rsstore )
% \]
% The \emph{extraction of dependence information} is a function
% $\taudv: \Dep \rightarrow \partsof{\Var}$ that is sound in the
% following sense:
% \[
% \forall \dep \in \Dep, \;
% \forall (\mem_0, \mem_1) \in \gammadep( \dep ), \;
% \mem_0 =_{\taudv( \dep )} \mem_1
% \]
\begin{definition}[Information translation and dependence abstraction]
  \label{d:8:pdep}
  The \emph{translation from symbolic to dependence} is a function
  $\tausd: \DSMem \rightarrow \Dep$ that is sound in the following
  sense:
  % \[
  $
  \forall \rsstore \in \DSMem, \;
  \forall (\mem_0, \mem_1, \valua) \in \dgammam( \rsstore ), \;
  (\mem_0, \mem_1) \in \gammadep \circ \tausd( \rsstore )
  $.
  % \]
  The \emph{extraction of dependence information} is a function
  $\taudv: \Dep \rightarrow \partsof{\Var}$ that is sound in the
  following sense:
  % \[
  $
  \forall \dep \in \Dep, \;
  \forall (\mem_0, \mem_1) \in \gammadep( \dep ), \;
  \mem_0 =_{\taudv( \dep )} \mem_1
  $
  % \]
\end{definition}
Intuitively, $\tausd$ should compute a dependence abstract domain
element that expresses a property implied by the relational symbolic
store it is applied to.
In the set-up of \eref{8:dep}, a straightforward way to achieve that
is to map $\rsstore$ to an element $\dep$ that maps $\Low$ to the set:
% \[
$
\{ \varx \in \Var \mid \; \;
% \vDash \projl( \rsstore )( \varx ) = \projr( \rsstore )( \varx ) \}
\may{\projl( \rsstore )( \varx ) = \projr( \rsstore )( \varx )}
\}
$
% \]

When $\rsstore( \varx ) = \langle \symex \rangle$, this equality is
clearly satisfied; when $\rsstore( \varx ) = \langle \symex_0 \mid
\symex_1 \rangle$, the equality $\symex_0 = \symex_1$ needs to be
discharged by an external tool such as an SMT solver.
Similarly, the function $\taudv$ extracts a set of variables which
are proved to remain low by the its argument.
In the setup of \eref{8:dep}, this boils down to returning $\dep(\Low)$.

We now present the combined analysis.
The symbolic execution step \textsc{\SRloopmanydep} is shown in
\fref{8:sdep} and replaces rule \textsc{\SRloopmany} (\fref{7:dsymstep}).
When the execution bound is reached for a loop statement, it performs
the dependence analysis of the whole loop from the dependence state
derived by applying $\tausd$ to the relational symbolic store.
Then, it applies $\taudv$ to derive the set of variables that are
proved to be low by the dependence analysis.
Finally, it computes a new relational symbolic store by modifying the
variables according to the set of variables determined low:
\begin{compactitem}
\item if variable $\varx$ is low based on the $\taudv$ output,
  $\sstatemodifydep$ synthesizes one
  fresh symbolic value $\syvarx_{\rm new}$ and maps it to $\msingle{\syvarx_{\rm new}}$;
\item if variable $\varx$ cannot be proved low, $\sstatemodifydep$
  synthesizes two fresh symbolic values $\syvarx_{{\rm new}0}$,
  $\syvarx_{{\rm new}1}$ and maps $\varx$ to $\mpair{\syvarx_{{\rm new}0}}{\syvarx_{{\rm new}1}}$.
\end{compactitem}
% \erasable{ The other rules of \fref{7:dsymstep} are not modified.}
%% TODO: simpler definition of modify: have two arguments low/high
\begin{figure}[t]
\[
\inferrule*[leftstyle={\sc},left=\SRloopmanydep]{
  \ctrstep(\stmtwhile{\prbcond}{\prcom},
  (\stmtseq{\prcom}{\stmtwhile{\prbcond}{\prcom}}), \counter) =
  (\false, \counter') \\
  \dep = \asemd{\stmtwhile{\prbcond}{\prcom}}(
  \tausd( \rsstore ) ) \\
  \rsstore'' = \sstatemodifydep( \rsstore, \prcom, \taudv( \dep ) ) \\
  (\prbcond, \rsstore'') \sreval \langle \beta_0, \beta_1 \rangle \\
  \spath' \triangleq \spath \logand \neg\beta_0 \logand \neg\beta_1
}{
  (\stmtwhile{\prbcond}{\prcom},
  (\rsstore, \spath), \counter, b)
  \rdse
  (\kwskip, (\rsstore'', \spath'), \counter', \false)
}
\]
  \caption{\RedSoundRSE: Symbolic execution approximation and product with dependence
    information.}
  \label{f:8:sdep}
\end{figure}
\begin{remark}[Reduced product property]
  We stress the fact that the rule {\sc\SRapproxmanydep} may be applied
  multiple times during the analysis, essentially whenever a loop
  statement is analyzed, which is generally many times more than the
  number of loop commands in the program due to abstract iterations.
  Therefore, our analysis \emph{cannot} be viewed as a fixed sequence
  of analyses.
  Such a decomposition (e.g., where dependence analysis is ran first
  and SE second) would be strictly less precise than our reduced
  product based approach.
\end{remark}

\paragraph{Soundness and refutation properties.}
Under the assumption that the dependence analysis and translation
operations are sound, so is the combined symbolic execution, thus
\tref{4:sounddsym} still holds.
Moreover, the refutation property of \tref{5:refute} also holds.
\begin{example}[Combined analysis]
  We consider program \ref{fp:b}.
  As discussed in \sref{2:over}, the loop statement may execute
  unboundedly many times, thus relational symbolic execution applies
  rule \textsc{\SRloopmanydep}.
  The initial dependence abstract element computed for the loop by
  $\tausd$ maps $\Low$ to $\{ \vari, \varz \}$ and $\High$ to all
  variables.
  The dependence analysis of the loop returns the same element.
  Thus, the set of low variables returned by $\taudv$ is $\{ \vari,
  \varz \}$, which allows to compute a precise relational symbolic store
  and to successfully verify the program is noninterferent.
\end{example}

%\section{Implementation and evaluation}
\section{Comparison} 
\label{section:prototype} \label{s:8:eval}

In this section we compare our analyses among them as well as with
the dependency analysis of Assaf et al.~\cite{anst:popl:17}.
To do so, we implemented prototypes of all the analyses. 
Our goal is not to evaluate the analyses in large code bases but to
assess their differences based on programs that are small but challenging
for typical noninteference analysers.

\paragraph{Implementation.}
We prototype the analyses proposed in this work as well as
the dependency analysis, intervals and convex polyhedra analysis.
The prototype is implemented in around 4k lines of OCaml code, using the  Apron library~\cite{jb:cav:09} for the numerical domains and the Z3 SMT solver~\cite{ldm:tacas:08}.
%Counters are implemented as stacks, where incrementing the counter means adding 1 to the top value
%of the stack, and exiting a loop is equivalent to popping the stack.
%The bound of iterations is constant in our implementation.
%We also provide an implementation for the function that calculates the set of modifiable variables.
%This set is over approximated in two steps: first it marks variables that appear on the left side on an
%assignment. Then, a second pass over the command is done where the value of variables marked
%in the first pass is cleared. This second pass makes use of the SMT solver to check if conditional
%statements can be accessed. It is important to have a decent over approximation of the set of
%modifiable variables to be able to maximize the utility of the dependence analysis.
By defining a shared interface for \SoundSE and \RedSoundSE, the implementation of \RedSoundRSE
is parameterized by these.
% \erasable{
% This allows us to switch between the standard \SSym and the reduced product of \SSym and a numerical
% analysis.
% To switch between the standard \DSym and the reduced product with dependences there is a boolean
% switch, which is optimal for code reuse since the rest of the rules are shared.
% }
An artifact of the implementation has been provided.

\begin{figure}[t]
  \resizebox{\textwidth}{!}{
    \subfigure[Secure]{ \label{fp:e}
\begin{lstlisting}[numbers=left]
if (priv > 0)
  i = 0;
else
  i = 0;
while (i < 10) {
  i += 1;
  priv += 5;
}
\end{lstlisting}}
    \qquad\quad
    % \subfigure[Secure]{\label{fp:f}
    %   \lstinputlisting[numbers=left]{prog-f-minimal.c}}
    % \qquad\quad
    \subfigure[Secure]{\label{fp:g}
\begin{lstlisting}[numbers=left]
i = 0; w = 2;
x = 100;
while(i < x) {
  if (x <= 0)
    w = priv;
  i += 2;
  x += 1;
}
\end{lstlisting}}
    \qquad\quad
    \subfigure[Insecure]{\label{fp:h}
\begin{lstlisting}[numbers=left]
i = 0;
while (i < 3) {
    y0 = y1;
    y1 = y2;
    y2 = priv;
    i += 1;
}
y1 = 0; y2 = 0;

\end{lstlisting}}
    \qquad\quad
    \subfigure[Insecure]{ \label{fp:i}
\begin{lstlisting}[numbers=left]
i = 0;
while (i < 100) {
    if (priv > 0)
        y = 5;
    i += 1;
}
\end{lstlisting}}
  }
  \caption{Programs illustrating different properties of the analyzer. Variable \varpriv\ is high.}
  \label{lst:imp:secure}
\end{figure}

\begin{table}[t]
\centering
\resizebox{\textwidth}{!}{%
\begin{tabular}{lc l ll ll}
\toprule
\multicolumn{2}{r}{\bf Relational Analysis}           & \multicolumn{1}{c}{$\Dep$} & \multicolumn{2}{c}{\SoundRSE}                                   & \multicolumn{2}{c}{\RedSoundRSE$(\Dep)$}                                \\ \cmidrule(lr){3-3} \cmidrule(lr){4-5} \cmidrule(lr){6-7}
\multicolumn{2}{r}{\bf relational analysis input:} & \multicolumn{1}{c}{None}   & \multicolumn{1}{c}{\SoundSE}  & \multicolumn{1}{c}{\RedSoundSE} & \multicolumn{1}{c}{\SoundSE} & \multicolumn{1}{c}{\RedSoundSE} \\ \cmidrule(r){1-2} \cmidrule(lr){3-3} \cmidrule(lr){4-5} \cmidrule(lr){6-7}
{\bf Program}             & {\bf Secure?} &                     &                               &                                   &                             &                                   \\ % \cmidrule(r){1-2} % \cmidrule(lr){3-3} \cmidrule(lr){4-5} \cmidrule(lr){6-7}
Fig.~\ref{fp:a}           & Yes           & \xmark\ False alarm & \cmark\ Secure                & \cmark\ Secure {\bf (I,P)}        & \cmark\ Secure              & \cmark\ Secure {\bf (I,P)}        \\ \midrule
Fig.~\ref{fp:b}           & Yes           & \cmark\ Secure      & \xmark\ False alarm           & \cmark\ Secure {\bf (P)}          & \cmark\ Secure              & \cmark\ Secure {\bf (I,P)}        \\ \midrule
Fig.~\ref{f:3:ex}         & Yes           & \xmark\ False alarm & \xmark\ False alarm           & \cmark\ Secure {\bf (I,P)}        & \xmark\ False alarm         & \cmark\ Secure {\bf (I,P)}        \\ \midrule
Fig.~\ref{fp:e}           & Yes           & \xmark\ False alarm & \xmark\ False alarm           & \xmark\ False alarm               & \cmark\  Secure             & \cmark\ Secure {\bf (I,P)}        \\ \midrule
Fig.~\ref{fp:g}           & Yes           & \xmark\ False alarm & \xmark\ False alarm           & \xmark\ False alarm               & \xmark\ False alarm         & \cmark\ Secure {\bf (I,P)}        \\ \midrule\midrule
Fig.~\ref{fp:c}           & No            & \cmark\ Alarm       & \cmark\ Refutation model      & \cmark\ Refutation model          & \cmark\ Refutation model    & \cmark\ Refutation model          \\ \midrule
Fig.~\ref{fp:h}           & No            & \cmark\ Alarm       & \cmark\ Refutation model      & \cmark\ Refutation model          & \cmark\ Refutation model    & \cmark\ Refutation model          \\ \midrule
Fig.~\ref{fp:i}           & No            & \cmark\ Alarm       & \cmark\ Alarm                 & \cmark\ Alarm                     & \cmark\ Alarm               & \cmark\ Alarm                     \\ \bottomrule
\end{tabular}}
\vskip 1em
\caption{Evaluation and comparison of analyses combination. $\Dep$ denotes the dependency analysis of ~\cite{anst:popl:17}.  Symbol \cmark\ (resp., \xmark\ ) denotes a semantically correct (resp.,
    incorrect) analysis outcome, with either a proof of security, a
    (possibly false) alarm, or a refutation model.
    % For examples using \RedSoundSE, we mark them with {\bf I} or {\bf P} to indicate that the
    % intervals domain or convex polyehdra domain are being used, respectively, for the product and
    % are succesful to prove N.I.}
    \modification{%
    For \RedSoundSE columns, when the analyses succeed to prove NI, we mark the result with {\bf I}
    (resp. {\bf P}) to indicate that the intervals (resp. polyhedra) domain is being used.}
    }
\label{f:10}
\vspace{-1.5em}
\end{table}

\paragraph{Evaluation.} We compare the 3 different relational techniques using different
single-trace analyses by evaluating them on a set of challenging examples. Our results are shown in
Table~\ref{f:10}. In the following, we split NI programs from non NI ones.
For the latter we look at the refutation capabilities of the analysis.
% \erasable{
% we first explain the verification capabilities of the different
% techniques for the secure examples and then their refutation capabilities for the insecure programs.
% }

\paragraph{Comparison of the verification capabilities of different relational analyses.}

Programs of Fig.~\ref{f:1:ex} were already explained in Section~\ref{s:2:over} and our prototype
confirmed these results, which are summarized in Table~\ref{f:10}.
%, is secure because it sets \vary\ to 5 in every execution. However, the dependence analysis does not take  values into account, therefore
%it raises a false alarm, as shown in the third column of Fig.~\ref{f:10}. Instead, 
% all of our relational symbolic execution-based analyses  can explore all paths and determine that the program
%is secure.
%In Program of Fig.~\ref{fp:b}, the dependence analysis can track \vari\ and determine there is no flow of information.
%Meanwhile, \SoundRSE over-approximates modified variables and loses all information of \vari.
% \erasable{
% We remark that, for Program of Fig.~\ref{fp:b}, the reduction of \SoundRSE and the convex polyhedra numerical abstraction
% proves the program secure because it can determine that the value of \vari\ is equal to the value of \varz. %intervals  can determine that \vari\ converges to 0.
% }

% \erasable{
% Program of Fig.~\ref{f:3:ex} can only be verified by using \RedSoundRSE instantiated with
% \RedSoundSE.
% The first $\kwif$ statement determines that $\varpriv \geq 0$.
% When the loop is over-approximated, numerical abstractions determine that the value of \vari\ is 10,
% and the states keep the constraint about \varpriv\ being positive.
% Hence, the last $\kwif$ cannot go through the false branch, and the value of \vary\ is always its
% original value plus one.
% }

In Program~\ref{fp:e}, the first condition renders dependence analysis
useless as it will consider variable \vari\ high.
This program will also fail to be verified by \SoundRSE if the iteration bound
is lower than 10: in this case, \vari\ will be assigned a fresh symbolic value
and hence be deemed high.
In contrast, \RedSoundRSE can determine that the value of \vari\ in the
loop does not depend on \varpriv.
%By using numerical domains, a false alarm is raised except if the user manually specifies a threshold
%for the widening of the loop.

% \erasable{
% Program~\ref{fp:f} has a loop with a high guard where the value of \vari\ is modified, therefore
% dependences is of no use.
% Likewise,  \SoundRSE instantiated with \SoundSE over-approximates and holds only the negation of the guards
% in the symbolic path, but this is not enough to determine that $\vari = 0$.
% In contrast, \SoundRSE instantiated with \RedSoundSE can determine the exact value of \vari\ and
% deem the program secure.
% }
% \modification{
Program~\ref{fp:g} is more convoluted. The analysis requires both numerical and dependence
abstractions in order to prove its NI.
The analysis will determine (conservatively) that three variables are modified in the loop: \varx,
\vari\ and \varw.
Dependence analysis can determine that variable \vari\ and \varx\ are low even if both are
modified.
However, since \varw\ depends on \varx, and the exact value of \varx\ is unknown, it is not possible
to determine that \varw\ is low.
By adding a numerical domain, it is easy to track that the value of \varx\ is always positive, which
implies that the $\kwif$ statement can never be executed.
% }

% Program~\ref{fp:g} TODO TALK WITH TAMARA AND XAVIER.

\paragraph{Comparison of the refutation capabilities of different relational analyses.}
Since \SoundRSE and \RedSoundRSE unroll loops a bounded number of times, there are insecure programs for
which a refutation model can be found, and programs where this is not possible.
Notice that, to refute a program with a model, it is required that
the symbolic execution did not
perform any over approximation, i.e. that the precision flag is set to false when the analysis finds the violation. Therefore, the results for insecure programs of \SoundRSE are similar to those
of the different combinations that rely on symbolic execution, as reflected on Figure~\ref{f:10}.
For Program~\ref{fp:c}, a valuation  can be found by doing one iteration: $\valua(\syvari_0) = \valua(\syvari_1) = 1$ and
$\valua(\syvarpriv_0) = 0$, $\valua(\syvarpriv_1) = 1$.
For Program~\ref{fp:h}, a model can be found if the bound of iterations is set to 4 or
higher.
The valuation $\valua$  just needs to map variable \varpriv\ to two different values:
$\valua(\syvarpriv_0) \not= \valua(\syvarpriv_1)$. 
In Program~\ref{fp:i}, for any user-set bound lower than 100 the execution will have to
overapproximate, losing refutation capabilities.

\paragraph{Conclusion of the evaluation.} We have evaluated and compared our analyses among them and
with the state-of-the-art on dependency analyses~\cite{anst:popl:17}  on a set of 8 challenging
examples. Our results show that, in contrast to dependencies~\cite{anst:popl:17},
analyses inherit the capacity of providing a refutation model  up to a bound from symbolic
execution. Moreover, \RedSoundRSE instantiated with \RedSoundSE is capable of soundly verifying all
the examples, in contrast to all the other compared analyses, as summarized in Table~\ref{f:10}.  

\paragraph{Limitations.}
As \RedSoundSE is sound and automatic, it necessarily fails to achieve
completeness (by Rice's Theorem~\cite{DBLP:books/daglib/0066920,DBLP:journals/jfrea/AspertiA08}).
% \erasable{
% , which means that there exist secure input programs that
% cannot be proven so.
% Obvious reasons for that are the approximations performed in the
% dependence abstraction and in the state abstraction.
% }
In return, we provide completeness up to a bound.
% \erasable{
% Our analysis is built so that the abstractions used are parameters and
% can be modified or switched, for instance by switching from interval abstraction
% to convex polyhedra, yet each abstract domain has limited expressiveness.
% Likewise, the computation of the variables that may be modified in a
% loop \erasable{(rule \textsc{s-a-many})} is also over-approximated.
% }
Another more subtle limitation is that the numerical abstraction are
applied at the level of the single symbolic execution (\RedSoundSE).
\modification{%
This means that these abstractions cannot track down relations between executions, but just local
constraints.
}%

\section{Related work} \label{s:9:rwc}
\paragraph{Hyperproperties} Noninterference was first defined by Goguen and Meseguer~\cite{gm:sp:82}, and also generalized to more powerful attacker models under the property name of declassification. 
We refer the reader to a survey on declassification policies~\cite{ss:ieee:05} up to 2005. 
As discussed in the introduction, noninterference is not a safety property but a safety hyperproperty~\cite{cs:hyper:08}, a.k.a. hypersafety. Several works in the literature have shown that hypersafety verification can be reduced to verification of safety properties~\cite{bar:csf:04,dhs:spc:05,ta:sas:05,cs:hyper:08}, however this reduction is not always efficient in practice~\cite{ta:sas:05}. In our work, we do not reduce noninterference to verification of safety but rather apply relational analyses. We only show our results using noninterference but the methodology can be easily generalized to more relaxed declassification properties, provided sound abstract domains exist.

\paragraph{Symbolic execution.} SE is a static analysis technique that was born in the 70s~\cite{bel:acm:75,king:acm:76} and that is now deployed in several popular testing tools, such as KLEE~\cite{cn:sttt:21} and
NASA's Symbolic PathFinder~\cite{pmbblpp:issta:08}, to name a few.
A primary goal and strength of SE is to find paths leading to counter-examples to generate  concrete input values exercising that
path. This is of particular importance to security in order to debug and confirm the feasibility of an attack when a vulnerability is detected. 

Alatawi et al.~\cite{DBLP:conf/kbse/AlatawiSM17} use AI to enhance
the precision of a dynamic symbolic execution aimed at path coverage.
Their approach consists of first doing an analysis of the program with AI to capture indirect
dependences in order to enhance path predicates.
%  on first analyzing the
% program with AI to capture the indirect dependencies on the outputs of the program,
% and then enhancing path predicates that would otherwise lead to path explosion.
% While this work also combines SE and AI, they alternate only once.
Furthermore, their analysis does not maintain
soundness (nor completeness).
Meanwhile, our approach continuously alternates between abstract domains and symbolic execution,
keeping soundness and completeness up to a bound.
Lastly, Alatawi et
al.~\cite{DBLP:conf/kbse/AlatawiSM17} do not analyze relational properties such as noninterference
but just safety properties.
% the aim is preventing path explosion, and
% they do not mantain soundness (nor completeness), while in our work we use AI to
% achieve soudness (without losing completeness up to a bound). Moreover, Alatawi et
% al.~\cite{DBLP:conf/kbse/AlatawiSM17} do not analyze relational properties such as noninterference
% but just safety properties.
% While, this might appear similar to our work there are two main differences:
% first, their proposed analysis is not sound nor complete, and lastly, it does not analyze
%relational
% properties such as noninterference but just safety properties.}
% Such an analysis cannot be sound nor complete.
% The analysis they propose focuses on path coverage in order to find bugs.
% The main difference lies on the goal of analysis. Their proposed analysis is not sound nor complete,
% and does not focus on relational properties such as noninterference.}
% The analysis they propose is not sound nor complete
% The combination done in the work of Alatawi et al. combines abstract interpretation and symbolic execution, but it
% does so sequentially while our framework supports a more synergetic way where both analysis communicate to each other.
% Lastly, the analyzer presented is not sound nor complete, and does not analyze noninterference or any hyperproperty.}

We focus the rest of the related work on static analysis techniques for relational security properties: 
for a broader discussion on symbolic execution  we refer the interested reader to a survey~\cite{cgkpstv:acm:11} up to 2011 and an illuminating discussion on SE challenges in practice up to 2013~\cite{cs:acm:13}.
\paragraph{Relational symbolic execution.} In order to apply SE to security properties such as noninterference, 
Milushev et al.~\cite{mbc:forte:12}  propose a form of relational symbolic execution (RSE) to use KLEE 
to analyze noninterference  
  by means of a technique called  self-composition~\cite{bar:csf:04,dhs:spc:05,ta:sas:05} to reduce 
a relational property of a program {\sf p} to a safety property of a transformation of {\sf p}.
More recently, Daniel et al. have optimized RSE to be applicable to binary code  to analyze relational properties such as constant time~\cite{dbt:sp:20} and speculative constant time~\cite{dbt:ndss:21,dbt:laser:22} and discovered violations of these properties in real-world cryptographic libraries.
All these approaches are based on pure (relational) SE static techniques and, as such, they are not 
capable of recovering soundness beyond a fixed bound as in our case.
The closest work to \DSym is  RelSym~\cite{fcg:ppdp:19}
which supports interactive refutation, as well as soundness.
In order to recover soundness, Chong et al.~\cite{fcg:ppdp:19}  propose to use RelSym on manually annotated programs with loop invariants. Precision of refutation is  guaranteed only if the invariants are strong enough, which cannot be determined by the tool itself. Precision is not guaranteed in any other cases.
In contrast, our invariants are automatically generated via AI and precision of
refutation is always guaranteed up to a bound, which is automatically computed by our tool.

\paragraph{Sound static analyses for hyperproperties} As discussed in the introduction, many sound verification methods have been proposed for relational security properties. We refer the reader to 
an excellent survey on this topic~\cite{sm:ieee:03} up to 2003. 
After 2003, several sound (semi-) static verification methods of noninterference-like properties
have been proposed by means of type systems (e.g.~\cite{bnr:sp:08,fpr:ccs:11}), hybrid types,
(e.g.~\cite{fjrs:tgc:15}), relational logics (e.g.~\cite{abggs:icfp:17}), model checking
(e.g.~\cite{hws:csf:06,bkr:sp:09}), and  pure AI~\cite{anst:popl:17}. 
We expand on the ones based on AI since they are the closest to our work. 
Giacobazzi and Mastroeni~\cite{gm:popl:04} define abstractions for attacker's views of program secrets and design 
sound automatic program analyses  based on AI for sets of executions (in contrast to relational executions).
Assaf et al.~\cite{anst:popl:17} are the first to express hyperproperties entirely within the 
framework of AI by defining a Galois connection that directly approximates the hyperproperty of
interest. We utilize the abstract domain of Assaf et al.~\cite{anst:popl:17} combined with SE to
obtain \RedSoundSE. 
Notice that because the framework of Assaf et al.~\cite{anst:popl:17}
relies on incomplete abstraction, their analysis is not
capable of precise refutation nor provide refutations models. 
To the best of our knowledge, no previous work has combined abstract domains and SE 
to achieve soundness. 
%npr:sp:18, impossiblitiyty of sound precise ts 

\section{Conclusion}\label{s:10:conc}
In this work, we propose a series of analyses, summarized in Fig.\ref{fig:symbolic_exec}, combining
SE and AI.
Our analyses are  sound, precise, and able to synthesize counter-examples  up
  to a given bound.
We prototype these analyses as well as several AI domains and a dependency analysis  to verify noninterference.
Our results, summarized in Table~\ref{f:10}, show that on a set of
challenging  examples for noninterference, our analysis performs better
than the dependency analysis and is able to preciselyblank and soundly
conclude on whether programs are noninterferent or not and provide
refutation models up to a bound. 
Given these encouraging  results, we plan to generalize the target security
property and make the analyses scale to other languages as future work.

\subsubsection*{Acknowledgements}
The authors thank the anonymous reviewers for
their comments, helpful for improving the paper. 
This project was funded by INRIA Challenge SPAI and by the VeriAMOS
ANR Project.
This research was partially supported by the ANR17-
CE25-0014-01 CISC project
We would also like to thank Josselin Giet and Adam Khayam for their observations.

\bibliographystyle{abbrv}
\bibliography{vmcai}

\newpage
\appendix

\modification{%
\section{Trace of program \ref{f:3:ex} with \RedSoundSE using intervals}\label{app:example}
}
This section aims to show the execution of one symbolic trace of program \ref{f:3:ex}.
Initial precise store $\sstate$ will capture the initial low-equality of variables $\vari$ and $\vary$.
The abstract state is $\astate$.
Changes to the product store are marked in \textred{red}.
\begin{equation*}
    \sstate =
      \begin{cases}
        \sstore = [ \vari \rightarrow \msingle{\syvari_0}, \vary \rightarrow \msingle{\syvary_0},
                    \varpriv \rightarrow \msingle{\syvarpriv_0}]\\
        \spath = \true \\
        \astate_l = [\ ]
      \end{cases}
\end{equation*}
In line 3, since $\varpriv$ is unconstrained, the semantics can choose either path.
Let us assume that our trace follows rule {\sc s-if-t}. Then, by line 5 the state is as follows.
\begin{equation*}
    \sstate =
      \begin{cases}
        \sstore = [ \vari \rightarrow \msingle{\syvari_0}, \vary \rightarrow \msingle{\syvary_0},
                    \varpriv \rightarrow \textred{\msingle{0}}]\\
        \spath = \textred{\syvarpriv_0 < 0} \\
        \astate = [ \textred{\varpriv = 0} ]
      \end{cases}
\end{equation*}
Since this loop has an unbounded amount of iterations, we know that an over approximation will happen.
Let us assume that the iteration bound is 1 (meaning that the semantics will execute
the loop once at most before over approximating), and that $\vari < 10$.
By executing one full iteration the following symbolic state is reached.
\begin{equation*}
    \sstate =
      \begin{cases}
        \sstore = [ \vari \rightarrow \textred{\msingle{\syvari_0 + 1}}, \vary \rightarrow \msingle{\syvary_0},
                    \varpriv \rightarrow \textred{\msingle{2}}]\\
        \spath = \textred{\syvari_0 < 10} \wedge \syvarpriv_0 < 0 \\
        \astate = [ \varpriv = 0 ; \textred{\vari < 11}] \qquad
      \end{cases}
\end{equation*}
Since now the limit of iterations is reached, next step is over approximating the loop.
For the example we will next show the state just before the reduction.
Notice that the new constraints in $\spath$ are the result of negating the guard.
\begin{equation*}
    \sstate =
      \begin{cases}
        \sstore = [ \vari \rightarrow \textred{\msingle{\syvari_1}}, \vary \rightarrow \msingle{\syvary_0},
                    \varpriv \rightarrow \textred{\msingle{\syvarpriv_1}}]\\
        \spath = \textred{\syvari_1 \geq 10} \wedge \syvari_0 < 10 \wedge \syvarpriv_0 < 0 \\
        \astate = [ \textred{\varpriv \geq 2} ; \textred{\vari = 10} ]
      \end{cases}
\end{equation*}
% \rho  = [ i \rightarrow < i2 | i3 > ; priv \rightarrow < p2 | p3 > ; y \rightarrow < y0 > ]
% \pi = i2 >= 10 \wedge i3 >= 10 \wedge (unimportant stuff)
% abstract state left = abstract state right = [ i = 10 ; priv >= 2 ]
%
Because variables $\vari$ and $\varpriv$ were modified, new symbolic values are assigned.
This generates a big inaccuracy, but abstract states can compensate.
By reducing we add the constraints of $\astate$ to $\spath$.
\begin{equation*}
    \sstate =
      \begin{cases}
        \sstore = [ \vari \rightarrow \msingle{\syvari_1}, \vary \rightarrow \msingle{\syvary_0},
                    \varpriv \rightarrow \msingle{\syvarpriv_1}]\\
        \spath = \textred{\syvarpriv_1 \geq 2 \wedge \syvari_1 = 10} \wedge \syvari_1 \geq 10 \wedge \syvari_0 < 10 \wedge \syvarpriv_0 < 0 \\
        \astate = [ \varpriv \geq 2 ; \vari = 10 ]
      \end{cases}
\end{equation*}
Thanks to the reduction, we get information allowing for the low equality of $\vari$ but also
we get information about $\varpriv$ being positive. Finally, the last $\kwif$ statement will not be
executed.

\section{SE step relation}\label{app:se}
This section shows the full set of rules of SE, the standard not-sound symbolic execution.
\[
\begin{array}{c}
  \inferrule*[leftstyle={\footnotesize \sc},left=s-assign]{
    (\prexpr, \sstore) \seval \symex
  }{
    (\varx := \prexpr, (\sstore, \spath))
    \sse
    (\kwskip, (\sstore[\varx \leadsto \msingle{\symex}], \spath))
  }
  \\[0.35ex]
  % S-SEQ-EXIT
  \inferrule*[leftstyle={\footnotesize \sc},left=s-seq-exit]{
  }{
    (\stmtseq{\stmtskip}{\prcom_1}, \sstate) \sse (\prcom_1, \sstate)
  }
  \quad
  % S-SEQ
  \inferrule*[leftstyle={\footnotesize \sc},left=s-seq]{
    (\prcom_0, \sstate) \sse (\prcom_0', \sstate')
  }{
    (\stmtseq{\prcom_0}{\prcom_1}, \sstate) \sse (\stmtseq{\prcom_0'}{\prcom_1}, \sstate')
  }
  \\[0.35ex]
  \inferrule*[leftstyle={\footnotesize \sc},left=s-if-t]{
    (\prbcond, \sstore) \seval \beta \\
    \spath' \triangleq \spath \wedge \beta \\
    \may{\spath'} \\
  }{
    (\stmtif{\prbcond}{\prcom_0}{\prcom_1},
    (\sstore, \spath))
    \sse
    (\prcom_0, (\sstore, \spath))
  }
  \\[0.35ex]
  \inferrule*[leftstyle={\footnotesize \sc},left=s-if-f]{
    (\prbcond, \sstore) \seval \beta \\
    \spath' \triangleq \spath \wedge \neg \beta \\
    \may{\spath'} \\
  }{
    (\stmtif{\prbcond}{\prcom_0}{\prcom_1},
    (\sstore, \spath))
    \sse
    (\prcom_1, (\sstore, \spath))
  }
  \\[0.35ex]
  \inferrule* [leftstyle={\footnotesize \sc},left=s-loop-t]{
    (\prbcond, \sstore) \seval \beta \\
    \spath' \triangleq \spath \wedge \beta \\
    \may{\spath'} \\
  }{
    (\stmtwhile{\prbcond}{\prcom}, (\sstore, \spath))
    \sse
    (\stmtseq{\prcom}{\stmtwhile{\prbcond}{\prcom}}, (\sstore, \spath))
  }
  \\[0.35ex]
  \inferrule* [leftstyle={\footnotesize \sc},left=s-loop-f]{
    (\prbcond, \sstore) \seval \beta \\
    \spath' \triangleq \spath \logand \neg\beta \\
    \may{\spath'} \\
  }{
    (\stmtwhile{\prbcond}{\prcom}, (\sstore, \spath))
    \sse
    (\kwskip, (\sstore, \spath))
  }
\end{array}
\]

\section{\SoundSE step relation}\label{app:soundse}
This section shows the full set of rules of \SoundSE by using SE, in Appendix~\ref{app:se}.
\[
\begin{array}{c}
  \inferrule*[leftstyle={\footnotesize \sc},left=s-next]{
    (\prcom, \sstate) \sse (\prcom', \sstate') \\
    \ctrstep(\prcom, \prcom', \counter) = (\true, \counter')
  }{
    (\prcom, \sstate, \counter, b)
    \sse
    (\prcom', \sstate', \counter', b) \\
  }
  \\[0.35ex]
  \inferrule*[leftstyle={\footnotesize \sc},left=s-approx-many]{
    (\prcom, \sstate) \sse (\prcom', \sstate') \\
    \ctrstep(\prcom, \prcom', \counter) = (\false, \counter') \\
    \sstore'' = \sstatemodify( \sstore, \prcom )
  }{
    (\prcom, (\sstore, \spath), \counter, b)
    \sse
    (\kwskip, (\sstore'', \spath), \counter', \false) \\
  }
\end{array}
\]

\newpage
\section{Abstract step relation}\label{app:abs}
This section shows the full set of rules of the abstract analysis used in \RedSoundSE.
\[
\begin{array}{c}
  \inferrule*[leftstyle={\footnotesize \sc},left=a-assign]{
    \astate' \triangleq \absassign_{\varx,\prexpr}( \astate )
  }{
    (\varx := \prexpr, \astate)
          \ssa
    (\kwskip, \astate')
  }
  \quad
  \inferrule*[leftstyle={\footnotesize \sc},left=a-if-t]{
    \astate' \triangleq \absguard_{\prbcond}( \astate ) \\
    \astate' \not= \bot
  }{
    (\stmtif{\prbcond}{\prcom_0}{\prcom_1},
    \astate)
          \ssa
    (\prcom_0, \astate')
  }
  \\[0.35ex]
  \inferrule*[leftstyle={\footnotesize \sc},left=a-if-f]{
    \astate' \triangleq \absguard_{\neg\prbcond}( \astate ) \\
    \astate' \not= \bot
  }{
    (\stmtif{\prbcond}{\prcom_0}{\prcom_1},
    \astate)
          \ssa
    (\prcom_0, \astate')
  }
  \\[0.35ex]
  % A-SEQ-EXIT
  \inferrule*[leftstyle={\footnotesize \sc},left=a-seq-exit]{
  }{
    (\stmtseq{\stmtskip}{\prcom_1}, \astate) \ssa (\prcom_1, \astate)
  }
  \quad
  % A-SEQ
  \inferrule*[leftstyle={\footnotesize \sc},left=a-seq]{
    (\prcom_0, \astate) \ssa (\prcom_0', \astate')
  }{
    (\stmtseq{\prcom_0}{\prcom_1}, \astate) \ssa (\stmtseq{\prcom_0'}{\prcom_1}, \astate')
  }
  \\[0.35ex]
  \inferrule*[leftstyle={\footnotesize \sc},left=a-loop-f]{
    \astate' \triangleq \absguard_{\neg\prbcond}( \astate ) \\
    \astate' \not= \bot
  }{
    (\stmtwhile{\prbcond}{\prcom_0}, \astate)
      \ssa
    (\stmtskip, \astate')
  }
  \\[0.35ex]
  \inferrule*[leftstyle={\footnotesize \sc},left=a-loop-t]{
    \astate' \triangleq \absguard_{\prbcond}( \astate ) \\
    \astate' \not= \bot
  }{
    (\stmtwhile{\prbcond}{\prcom_0}, \astate)
      \ssa
    (\stmtseq{\prcom}{\stmtwhile{\prbcond}{\prcom_0}}, \astate')
  }
  \\[0.35ex]
\end{array}
\]

\section{\RedSoundSE step relation}
\RedSoundSE is defined by rules of Appendix~\ref{app:se}, Appendix~\ref{app:soundse} and
Appendix~\ref{app:abs}.
\[
\begin{array}{c}
  \inferrule*[leftstyle={\footnotesize \sc},left=s-a-next]{
    (\prcom, \sstate, \counter, b)
    \sse
    (\prcom', \sstate', \counter', b) \\
    \ctrstep(\prcom, \prcom', \counter) = (\true, \counter') \\
    (\prcom, \astate) \ssa (\prcom', \astate') \\
    (\sstate'', \astate'') \triangleq \reduction( \sstate', \astate' )
  }{
    (\prcom, \sstate, \astate, \counter, b)
    \rsse
    (\prcom', \sstate'', \astate'', \counter', b) \\
  }
  \\[2ex]
  \inferrule*[leftstyle={\footnotesize \sc},left=s-a-approx-many]{
    (\prcom, \sstate) \sse (\prcom', \sstate') \\
    \ctrstep(\prcom, \prcom', \counter) = (\false, \counter') \\
    \sstate'' = \sstatemodify( \sstate, \prcom ) \\
    \astate' = \asema{\prcom}( \astate ) \\
    (\sstate''', \astate''') \triangleq \reduction( \sstate'', \astate')
  }{
    (\prcom, (\sstate, \astate), \counter, b)
    \rsse
    (\kwskip, (\sstate''', \astate'''), \counter', \false) \\
  }
  \\
\vspace{0.01em}
\end{array}
\]

\newpage

\section{RSE and \SoundRSE step relations}\label{app:soundrse}
This section shows the full set of rules for \SoundRSE. RSE is a subset of \SoundRSE, by removing
rule \textsc{\SRapproxmany}, and removing the counter and boolean flag.
\[
\begin{array}{c}
  \inferrule*[leftstyle={\footnotesize \sc},left=\SRassign]{
    (\prexpr, \rsstore) \sreval \rsymex
  }{
    (\varx := \prexpr, (\rsstore, \spath), \counter, b)
    \dse
    (\kwskip, (\rsstore[\varx \leadsto \langle \rsymex \rangle], \spath),
    \counter, b)
  }
  \\[0.75ex]
  % SEQ-EXIT
  \inferrule*[leftstyle={\footnotesize \sc},left=sr-seq-exit]{
  }{
    (\stmtseq{\stmtskip}{\prcom_1}, \rsstate) \dse (\prcom_1, \rsstate)
  }
  \quad
  % SEQ
  \inferrule*[leftstyle={\footnotesize \sc},left=sr-seq]{
    (\prcom_0, \rsstate) \dse (\prcom_0', \rsstate')
  }{
    (\stmtseq{\prcom_0}{\prcom_1}, \rsstate) \dse (\stmtseq{\prcom_0'}{\prcom_1}, \rsstate')
  }
  \\[0.75ex]
  % IF-TT
  \inferrule*[leftstyle={\footnotesize \sc},left=\SRiftt]{
    (\prbcond, \rsstore) \sreval \dbeta \\
    \spath' = \spath \logand \projl(\dbeta) \logand \projr(\dbeta) \\
    \may{\spath'} \\
  }{
    (\stmtif{\prbcond}{\prcom_0}{\prcom_1}, (\rsstore, \spath), \counter, b) \dse
    (\prcom_0, (\rsstore, \spath'), \counter, b)
  % \inferrule*[leftstyle={\footnotesize \sc},left=\SRiftt]{
  %   (\kwif \, \prbcond \, \kwthen \, \prcom_0 \, \kwelse \, \prcom_1,
  %   (\projl(\rsstore), \spath), \counter, b)
  %   \sse
  %   (\prcom_0, (\sstore'_0, \spath'_0), \counter, b) \\
  %   (\kwif \, \prbcond \, \kwthen \, \prcom_0 \, \kwelse \, \prcom_1,
  %   (\projr(\rsstore), \spath), \counter, b)
  %   \sse
  %   (\prcom_0, (\sstore'_1, \spath'_1), \counter, b) \\
  %   (\kwif \, \prbcond \, \kwthen \, \prcom_0 \, \kwelse \, \prcom_1,
  %   (\rsstore, \spath), \counter, b)
  %   \dse
  %   (\prcom_0, (\pairing{\sstore'_0}{\sstore'_1}, \spath'_0 \logand
  %   \spath'_1), \counter, b)
  }
  \\[0.75ex]
  % IF-TF
  \inferrule*[leftstyle={\footnotesize \sc},left=\SRiftf]{
    (\prbcond, \rsstore) \sreval \dbeta \\
    \spath' = \spath \logand \projl(\dbeta) \logand \neg \projr(\dbeta) \\
    \may{\spath'} \\
  }{
    (\stmtif{\prbcond}{\prcom_0}{\prcom_1}, (\rsstore, \spath), \counter, b) \dse
    (\prcom_0 \bowtie \prcom_1, (\rsstore, \spath'), \counter, b)
  }
  % \inferrule*[leftstyle={\footnotesize \sc},left=\SRiftf]{
  %   (\kwif \, \prbcond \, \kwthen \, \prcom_0 \, \kwelse \, \prcom_1,
  %   (\projl(\rsstore), \spath), \counter, b)
  %   \sse
  %   (\prcom_0, (\sstore'_0, \spath'_0), \counter, b) \\
  %   (\kwif \, \prbcond \, \kwthen \, \prcom_0 \, \kwelse \, \prcom_1,
  %   (\projr(\rsstore), \spath), \counter, b)
  %   \sse
  %   (\prcom_1, (\sstore'_1, \spath'_1), \counter, b) \\
  % }{
  %   (\kwif \, \prbcond \, \kwthen \, \prcom_0 \, \kwelse \, \prcom_1,
  %   (\rsstore, \spath), \counter, b)
  %   \dse
  %   (\prcom_0 \bowtie \prcom_1,
  %   (\pairing{\sstore'_0}{\sstore'_1}, \spath'_0 \logand \spath'_1),
  %   \counter, b)
  % }
  \\[0.75ex]
  % IF-FT
  \inferrule*[leftstyle={\footnotesize \sc},left=\SRifft]{
    (\prbcond, \rsstore) \sreval \dbeta \\
    \spath' = \spath \logand \neg \projl(\dbeta) \logand \projr(\dbeta) \\
    \may{\spath'} \\
  }{
    (\stmtif{\prbcond}{\prcom_0}{\prcom_1}, (\rsstore, \spath), \counter, b) \dse
    (\prcom_1 \bowtie \prcom_0, (\rsstore, \spath'), \counter, b)
  }
  % \inferrule*[leftstyle={\footnotesize \sc},left=\SRifft]{
  %   (\kwif \, \prbcond \, \kwthen \, \prcom_0 \, \kwelse \, \prcom_1,
  %   (\projl(\rsstore), \spath), \counter, b)
  %   \sse
  %   (\prcom_1, (\sstore'_0, \spath'_0), \counter, b) \\
  %   (\kwif \, \prbcond \, \kwthen \, \prcom_0 \, \kwelse \, \prcom_1,
  %   (\projr(\rsstore), \spath), \counter, b)
  %   \sse
  %   (\prcom_0, (\sstore'_1, \spath'_1), \counter, b) \\
  % }{
  %   (\kwif \, \prbcond \, \kwthen \, \prcom_0 \, \kwelse \, \prcom_1,
  %   (\rsstore, \spath), \counter, b)
  %   \dse
  %   (\prcom_1 \bowtie \prcom_0,
  %   (\pairing{\sstore'_0}{\sstore'_1}, \spath'_0 \logand \spath'_1),
  %   \counter, b)
  % }
  \\[0.75ex]
  % IF-FF
  \inferrule*[leftstyle={\footnotesize \sc},left=\SRifff]{
    (\prbcond, \rsstore) \sreval \dbeta \\
    \spath' = \spath \logand \neg \projl(\dbeta) \logand \neg \projr(\dbeta) \\
    \may{\spath'} \\
  }{
    (\stmtif{\prbcond}{\prcom_0}{\prcom_1}, (\rsstore, \spath), \counter, b) \dse
    (\prcom_0, (\rsstore, \spath'), \counter, b)
  }
  % \inferrule*[leftstyle={\footnotesize \sc},left=\SRifff]{
  %   (\kwif \, \prbcond \, \kwthen \, \prcom_0 \, \kwelse \, \prcom_1,
  %   (\projl(\rsstore), \spath), \counter, b)
  %   \sse
  %   (\prcom_1, (\sstore'_0, \spath'_0), \counter, b) \\
  %   (\kwif \, \prbcond \, \kwthen \, \prcom_0 \, \kwelse \, \prcom_1,
  %   (\projr(\rsstore), \spath), \counter, b)
  %   \dse
  %   (\prcom_1, (\sstore'_1, \spath'_1), \counter, b) \\
  % }{
  %   (\kwif \, \prbcond \, \kwthen \, \prcom_0 \, \kwelse \, \prcom_1,
  %   (\rsstore, \spath), \counter, b)
  %   \sse
  %   (\prcom_1, (\pairing{\sstore'_0}{\sstore'_1}, \spath'_0 \logand
  %   \spath'_1), \counter, b)
  % }
  \\[0.75ex]
  % LOOP-TT
  \inferrule*[leftstyle={\footnotesize \sc},left=\SRlooptt]{
    \ctrstep(\prcom, \prcom', \counter) = (\true, \counter') \\
    (\prbcond, \rsstore) \sreval \dbeta \\
    \spath' = \spath \logand \projl(\dbeta) \logand \projr(\dbeta) \\
    \may{\spath'} \\
    % (\stmtwhile{\prbcond}{\prcom_0}, (\projl(\rsstore), \spath), \counter, b) \sse
    %   (\prcom_0; \stmtwhile{\prbcond}{\prcom_0}, (\sstore'_0, \spath'_0), \counter', b) \\
    % (\stmtwhile{\prbcond}{\prcom_0}, (\projr(\rsstore), \spath), \counter, b) \sse
    %   (\prcom_0 ; \stmtwhile{\prbcond}{\prcom_0}, (\sstore'_1, \spath'_1), \counter', b) \\
  }{
    (\stmtwhile{\prbcond}{\prcom_0}, (\rsstore, \spath), \counter, b) \dse
    (\stmtseq{\prcom_0}{\stmtwhile{\prbcond}{\prcom_0}}, (\rsstore, \spath), \counter', b)
  }
  \\[0.75ex]
  % LOOP-TF
  \inferrule*[leftstyle={\footnotesize \sc},left=\SRlooptf]{
    \ctrstep(\prcom, \prcom', \counter) = (\true, \counter') \\
    (\prbcond, \rsstore) \sreval \dbeta \\
    \spath' = \spath \logand \projl(\dbeta) \logand \neg \projr(\dbeta) \\
    \may{\spath'} \\
  }{
    (\stmtwhile{\prbcond}{\prcom_0}, (\rsstore, \spath), \counter, b) \dse
      ((\stmtseq{\prcom_0}{\stmtwhile{\prbcond}{\prcom_0}}) \bowtie \stmtskip, (\rsstore, \spath'), \counter', b)
  }
  \\[0.75ex]
  % LOOP-FT
  \inferrule*[leftstyle={\footnotesize \sc},left=\SRloopft]{
    \ctrstep(\prcom, \prcom', \counter) = (\true, \counter') \\
    (\prbcond, \rsstore) \sreval \dbeta \\
    \spath' = \spath \logand \neg \projl(\dbeta) \logand \projr(\dbeta) \\
    \may{\spath'} \\
  }{
    (\stmtwhile{\prbcond}{\prcom_0}, (\rsstore, \spath), \counter, b) \dse
      (\stmtskip \bowtie (\stmtseq{\prcom_0}{\stmtwhile{\prbcond}{\prcom_0}}), (\rsstore, \spath'), \counter', b)
  }
  \\[0.75ex]
  % LOOP-FF
  \inferrule*[leftstyle={\footnotesize \sc},left=\SRloopff]{
    \ctrstep(\prcom, \prcom', \counter) = (\true, \counter') \\
    (\prbcond, \rsstore) \sreval \dbeta \\
    \spath' = \spath \logand \neg \projl(\dbeta) \logand \neg \projr(\dbeta) \\
    \may{\spath'} \\
  }{
    (\stmtwhile{\prbcond}{\prcom_0}, (\rsstore, \spath), \counter, b) \dse
    (\stmtskip, (\rsstore, \spath'), \counter', b)
  }
  \\[1.75ex]
  \inferrule*[leftstyle={\footnotesize \sc},left=\SRexit]{
  }{
    (\kwskip \bowtie \kwskip, (\rsstore, \spath), \counter, b)
    \dse
    (\kwskip, (\rsstore, \spath), \counter, b)
  }
  \\[0.75ex]
  \inferrule*[leftstyle={\footnotesize \sc},left=\SRcompr]{
    (\prcom_1, (\projr(\rsstore), \spath), \counter, b)
    \sse
    (\prcom'_1, (\sstore'_1, \spath'), \counter', b')
  }{
    (\kwskip \bowtie \prcom_1, (\rsstore, \spath), \counter, b)
    \dse
    (\kwskip \bowtie \prcom'_1, (\pairing{\projl(\rsstore)}{\sstore'_1},
    \spath'), \counter', b')
  }
  \\[0.75ex]
  \inferrule*[leftstyle={\footnotesize \sc},left=\SRcompl]{
    (\prcom_0, (\projl(\rsstore), \spath), \counter, b)
    \sse
    (\prcom'_0, (\sstore'_0, \spath'), \counter', b')
  }{
    (\prcom_0 \bowtie \prcom_1, (\rsstore, \spath), \counter, b)
    \dse
    (\prcom'_0 \bowtie \prcom_1, (\pairing{\sstore'_0}{\projr(\rsstore)},
    \spath'), \counter', b')
  }
  \\[0.75ex]
  \inferrule*[leftstyle={\footnotesize \sc},left=\SRloopmany]{
    \ctrstep(\stmtwhile{\prbcond}{\prcom},
    (\stmtseq{\prcom}{\stmtwhile{\prbcond}{\prcom}}), \counter) =
    (\false, \counter') \\
    \rsstore'' = \sstatemodify( \rsstore, \prcom ) \\
    (\prbcond, \rsstore'') \sreval \langle \beta_0, \beta_1 \rangle \\
    \spath' \triangleq \spath \logand \neg\beta_0 \logand \neg\beta_1
  }{
    (\stmtwhile{\prbcond}{\prcom},
    (\rsstore, \spath), \counter, b)
    \dse
    (\kwskip, (\rsstore'', \spath'), \counter', \false)
  }
  % \inferrule*[leftstyle={\footnotesize \sc},left=\SRloopmany]{
  %   \ctrstep(\stmtwhile{\prbcond}{\prcom},  \prcom ; \stmtwhile{\prbcond}{\prcom}, \counter) =
  %   (\false, \counter') \\
  %   \rsstore'' = \sstatemodify( \rsstore, \prcom ) \\
  %   (\prbcond, \rsstore'') \sreval \langle \beta_0, \beta_1 \rangle \\
  %   \spath' \triangleq \spath \logand \neg\beta_0 \logand \neg\beta_1
  % }{
  %   (\kwwhile \, \prbcond \, \kwdo \, \prcom,
  %   (\rsstore, \spath), \counter, b)
  %   \dse
  %   (\kwskip, (\rsstore'', \spath'), \counter', \false)
  % }
  \\
\end{array}
\]

\section{\RedSoundRSE step relation}
\RedSoundRSE is defined by rules of Appendix~\ref{app:soundrse} plus rule \textsc{\SRloopmanydep}.
\[
\inferrule*[leftstyle={\sc},left=\SRloopmanydep]{
  \ctrstep(\stmtwhile{\prbcond}{\prcom},
  (\stmtseq{\prcom}{\stmtwhile{\prbcond}{\prcom}}), \counter) =
  (\false, \counter') \\
  \dep = \asemd{\stmtwhile{\prbcond}{\prcom}}(
  \tausd( \rsstore ) ) \\
  \rsstore'' = \sstatemodifydep( \rsstore, \prcom, \taudv( \dep ) ) \\
  (\prbcond, \rsstore'') \sreval \langle \beta_0, \beta_1 \rangle \\
  \spath' \triangleq \spath \logand \neg\beta_0 \logand \neg\beta_1
}{
  (\stmtwhile{\prbcond}{\prcom},
  (\rsstore, \spath), \counter, b)
  \rdse
  (\kwskip, (\rsstore'', \spath'), \counter', \false)
}
\]

\end{document}